\documentclass[preprint]{aastex63}       

\usepackage{longtable} 
\usepackage{threeparttable} 
\usepackage{booktabs}    
\usepackage{lineno}
\usepackage{bibunits}
\usepackage{bm}
\usepackage{amsmath, amssymb}
\newcommand{\insight}{\textit{Insight}-HXMT}

\accepted{by ApJ}

\shorttitle{Fan beam emission and magnetic field of RX J0209.6$-$7427}
\shortauthors{Hou et al.}

\begin{document}

\title{Fan beamed X-ray emission from 1 keV to above 130 keV from the ultraluminous X-ray pulsar RX J0209.6$-$7427 in the Small Magellanic Cloud}

\correspondingauthor{X. Hou, S.N. Zhang}
\email{xhou@ynao.ac.cn, zhangsn@ihep.ac.cn}

\author[0000-0003-0933-6101]{X. Hou}
\affiliation{Yunnan Observatories, Chinese Academy of Sciences, Kunming 650216, China}
\affiliation{Key Laboratory for the Structure and Evolution of Celestial Objects, Chinese Academy of Sciences, Kunming 650216, China}

\author[0000-0002-2749-6638]{M.Y. Ge}
\affiliation{Key Laboratory for Particle Astrophysics, Institute of High Energy Physics, Chinese Academy of Sciences, Beijing 100049, China}

\author[0000-0001-9599-7285]{L. Ji}
\affiliation{School of Physics and Astronomy, Sun Yat-Sen University, Zhuhai 519082, China}

\author[0000-0001-5586-1017]{S.N. Zhang}
\affiliation{Key Laboratory for Particle Astrophysics, Institute of High Energy Physics, Chinese Academy of Sciences, Beijing 100049, China}
\affiliation{University of Chinese Academy of Sciences, Chinese Academy of Sciences, Beijing 100049, China}

\author{Y. You}
\affiliation{Key Laboratory for Particle Astrophysics, Institute of High Energy Physics, Chinese Academy of Sciences, Beijing 100049, China}
\affiliation{University of Chinese Academy of Sciences, Chinese Academy of Sciences, Beijing 100049, China}

\author[0000-0002-2705-4338]{L. Tao}
\affiliation{Key Laboratory for Particle Astrophysics, Institute of High Energy Physics, Chinese Academy of Sciences, Beijing 100049, China}

\author{S. Zhang}
\affiliation{Key Laboratory for Particle Astrophysics, Institute of High Energy Physics, Chinese Academy of Sciences, Beijing 100049, China}

\author[0000-0002-4622-796X]{R. Soria}
\affiliation{University of Chinese Academy of Sciences, Chinese Academy of Sciences, Beijing 100049, China}

\author[0000-0001-7584-6236]{H. Feng}
\affiliation{Department of Astronomy, Tsinghua University, Beijing 100084, China}

\author{M. Zhou}
\affiliation{Yunnan Observatories, Chinese Academy of Sciences, Kunming 650216, China}
\affiliation{Key Laboratory for the Structure and Evolution of Celestial Objects, Chinese Academy of Sciences, Kunming 650216, China}

\author{Y.L. Tuo}
\affiliation{Key Laboratory for Particle Astrophysics, Institute of High Energy Physics, Chinese Academy of Sciences, Beijing 100049, China}

\author{L.M. Song}
\affiliation{Key Laboratory for Particle Astrophysics, Institute of High Energy Physics, Chinese Academy of Sciences, Beijing 100049, China}
\affiliation{University of Chinese Academy of Sciences, Chinese Academy of Sciences, Beijing 100049, China}

\author{J.C. Wang}
\affiliation{Yunnan Observatories, Chinese Academy of Sciences, Kunming 650216, China}
\affiliation{Key Laboratory for the Structure and Evolution of Celestial Objects, Chinese Academy of Sciences, Kunming 650216, China}
\affiliation{University of Chinese Academy of Sciences, Chinese Academy of Sciences, Beijing 100049, China}

\begin{abstract}

We present detailed timing and spectral analyses of the transient X-ray pulsar RX J0209.6$-$7427 in the Small Magellanic Cloud during its 2019 giant outburst. With a better known distance than most galactic X-ray pulsars, its peak luminosity is determined to be $(1.11\pm0.06)\times 10^{39}\, \rm erg\ s^{-1}$; it is thus a {\it bonda fide} pulsating ultraluminous X-ray source (PULX). Owing to the broad energy band of \textit{Insight}-HXMT, its pulsed X-ray emission was detected from 1 keV up to the 130$-$180 keV band, which is the highest energy emission detected from any PULXs outside the Milky Way. This allows us to conclude that its main pulsed X-ray emission is from the ``fan beam'' of the accretion column, and its luminosity is thus intrinsic. We also estimate its magnetic field of (4.8$-$8.6)$\times10^{12}$ G or (1.7$-$2.2)$\times10^{13}$ G, from its spin evolution or transition in the accretion column structure during the outburst; we suggest that the two values of the magnetic field strength correspond to the dipole and multipole magnetic fields of the neutron star, similar to the recent discovery in the Galactic PULX Swift J0243.6+6124. Therefore, the nature of the neutron star and its ULX emission can be understood within the current theoretical frame of accreting neutron stars. This may have implications for understanding the nature of those farther away extragalactic PULXs.

\end{abstract}


\section{Introduction} \label{sec:intro}

Ultraluminous X-ray sources (ULXs) are objects first detected in nearby galaxies with apparent luminosity $\gtrsim 10^{39}\, \rm erg\ s^{-1}$, above the Eddington limit for a stellar-mass black hole (BH) ($\sim 10M_{\rm \odot}$) \citep{Kaaret2017}, but which are not supermassive BHs. The recent discoveries of coherent pulsations in ULXs unambiguously established that accreting neutron stars (NSs) can be the central engines of ULXs \citep{Bachetti2014,Furst2016,Israel2017a,Israel2017b,Carpano2018}. The radiative mechanisms of pulsating ULXs (PULXs) is still under debate. Different scenarios have been proposed to explain their super-Eddington luminosity, involving NSs with strong magnetic dipole fields or multipolar fields close to that of magnetars \citep{Mushtukov2015a,Chashkina2017,Israel2017a,Chashkina2019}, or NSs with standard magnetic fields ($10^{12}-10^{13}$ G) but whose emission is highly collimated rather than quasi-isotropic \citep{King2009,Kluzniak2015,King2016,Koliopanos2017,Pintore2017,King2017,Middleton2017,Walton2018,King2019,King2020}. Until recently, PULXs have mostly been found in extragalactic galaxies at a distance of a few Mpc. Therefore, detailed study on PULXs is hampered by the limited number of PULXs detected and limited X-ray observations.

RX J0209.6$-$7427 is a transient accreting X-ray pulsar \citep[see][for a recent review on X-ray pulsars]{Mushtukov2022} in the outer wing of the SMC discovered during its outburst in 1993 \citep{Kahabka2005}. During its recent 2019 giant outburst, a spin period of 9.2\,s and a peak luminosity of $\sim 10^{39} \, \rm erg\ s^{-1}$ in the energy range of 0.2$-$12 keV was first reported using \textit{NICER} data \citep{Iwakiri2019,Vasilopoulos2020}. Pulsations were then detected in 8$-$50 keV with the \textit{Fermi}/GBM data, in 3$-$70 keV with the \textit{NuSTAR} data \citep{Vasilopoulos2020} and in 3$-$80 keV with the \textit{AstroSat} data \citep{Chandra2020}. RX J0209.6$-$7427 is thus established as a new PULX. Optical observations classified its companion star to be a Be type \citep{Coe2020}. Compared to the extragalactic PULXs, this source has the advantage of proximity to allow detailed observational diagnoses. Compared to the first Galactic PULX Swift J0243.6+6124 \citep{Wilson-Hodge2018}, whose distance has rather large uncertainty ranging from 5 kpc to 8.9 kpc \citep{van2018}, RX J0209.6$-$7427 has a better determined distance of 55 kpc and much smaller relative distance uncertainty of 5\%\footnote{The mean distance modulus of the SMC is $18.89\pm0.04 \, \rm (stat) \pm0.10 \, \rm(syst)$, which corresponds to a distance of $60\pm 2.8 \, \rm (stat+syst)$ kpc \citep{Harries2003}. However, a smaller modulus of 18.7 corresponding to 55 kpc is usually adopted for sources located at the SMC wing, but no uncertainty was reported \citep{Cignoni2009}. We thus adopted the same uncertainty of 2.8 kpc as for the SMC mean distance.}, hence less uncertainty on its luminosity. Therefore, RX J0209.6$-$7427 is an ideal source to gain insight into the PULX emission properties.

In this work, we analyze \textit{Insight}-HXMT data and more \textit{NICER} data than previously reported \citep{Vasilopoulos2020} to characterize the spectral and temporal properties of RX J0209.6$-$7427, so as to investigate its ultraluminous emission origin. The paper is organized as follows: observations and data reduction method are described in Section \ref{obs}, and the detailed data analysis results are presented in Section \ref{result}. We discuss our results in Section \ref{discuss} and conclude in Section \ref{conclude}.

\section{Observations and data reduction} \label{obs}

\subsection{\textit{NICER}}
\textit{NICER}, launched on 2017 June 03, is an International Space Station payload devoted to the study of NSs through high sensitivity X-ray timing in the soft (0.2$-$12 keV) X-ray band \citep{Gendreau2016}. Its X-ray Timing Instrument (XTI) is an aligned collection of 56 X-ray concentrator optics (XRC) and silicon drift detector (SDD) pairs. Each XRC collects X-rays over a large geometric area from a roughly 30 arcmin$^{2}$ region of the sky and focuses them on to a small SDD. The SDD detects individual photons, recording their energies with good spectral resolution and their detection times to a $\sim$100 nanoseconds RMS relative to the Universal Time. 

We used \textit{NICER} observations from 2019 November 21 to  2020 March 20 (58808-58928 MJD) covering both the rising and decaying parts of the ourburst. \textit{NICER} data reduction is performed using \texttt{HEASOFT} (version 6.28). For the timing analysis, the good time intervals (GTIs) are selected according to the following criteria: \textit{NICER} not in the South Atlantic Anomaly region, source elevation $>10^{\circ}$ above the Earth limb ($>20^{\circ}$ above the bright Earth), pointing offset $\lesssim 54^{\prime}$, and magnetic cut-off rigidity $>1.5$ GeV/c. We further correct the arrival time of every event to barycentre via the \texttt{barycorr} tool and the JPL-DE405 planetary ephemeris.

\subsection{\textit{Insight}-HXMT}
Launched on 2017 June 15, the Hard X-ray Modulation Telescope \citep{Zhang2020} (\textit{Insight}-HXMT) is the first Chinese X-ray astronomy satellite. There are three scientific payloads onboard the satellite: the Low Energy X-ray telescope (LE, 1$-$10 keV) \citep{Chen2020}, the Medium Energy X-ray telescope (ME, 5$-$30 keV) \citep{Cao2020}, and the High Energy X-ray telescope (HE, 20$-$250 keV) \citep{Liu2020}. 

\textit{Insight}-HXMT observed RX J0209.6-7427 from 2019 December 10 to December 15 (58827-58832 MJD) around the outburst peak, for a total exposure of 100 ks. The data reduction is performed using the \textit{Insight}-HXMT data analysis software package \texttt{HXMTDAS} v2.04\footnote{http://www.hxmt.org/software.jhtml} following the standard procedure in the \texttt{HXMTDAS} user guide\footnote{http://www.hxmt.org/SoftDoc/67.jhtml}. Guide lines of the procedure is as following: (1) Use the commands \texttt{hepical}, \texttt{mepical} and \texttt{lepical} to calibrate the photon events from the raw data according to the Calibration Database (CALDB) of \textit{Insight}-HXMT. (2) Select the GTIs using the commands \texttt{hegtigen}, \texttt{megtigen} and \texttt{legtigen} for calibrated events. (3) Extract the good events basing on the GTIs using the commands \texttt{hescreen}, \texttt{mescreen} and \texttt{lescreen}. (4) Generate spectra for the selected events using the commands \texttt{hespecgen}, \texttt{mespecgen} and \texttt{lespecgen}. (5) Generate the background spectra basing on the emission detected by blind detectors using the commands \texttt{hebkgmap}, \texttt{mebkgmap} and \texttt{lebkgmap}. (6) Generate the response matrix files required for spectral analysis using the commands \texttt{herspgen}, \texttt{merspgen} and \texttt{lerspgen}.

In the timing analysis, the arrival times of all the cleaned events are further barycenter corrected via the \texttt{hxbary} tool and the JPL-DE405 planetary ephemeris. In the spectral analysis, only the small field of view (FOV) detectors from the LE and ME instruments are used. The spectral fit is performed using \texttt{XSPEC} v12.11.0 \citep{1996ASPC..101...17A}. Uuncertainties are reported at the 68\% confidence interval and are computed using Markov Chain Monte Carlo simulations (MCMC, available through \texttt{XSPEC}) of length $10^{5}$.

\section{Data analysis and results} \label{result}

\subsection{Long-term flux and spectral evolution}
\label{lcHID}
\begin{figure}[ht!]
	\centering
	\includegraphics[width=0.5\linewidth]{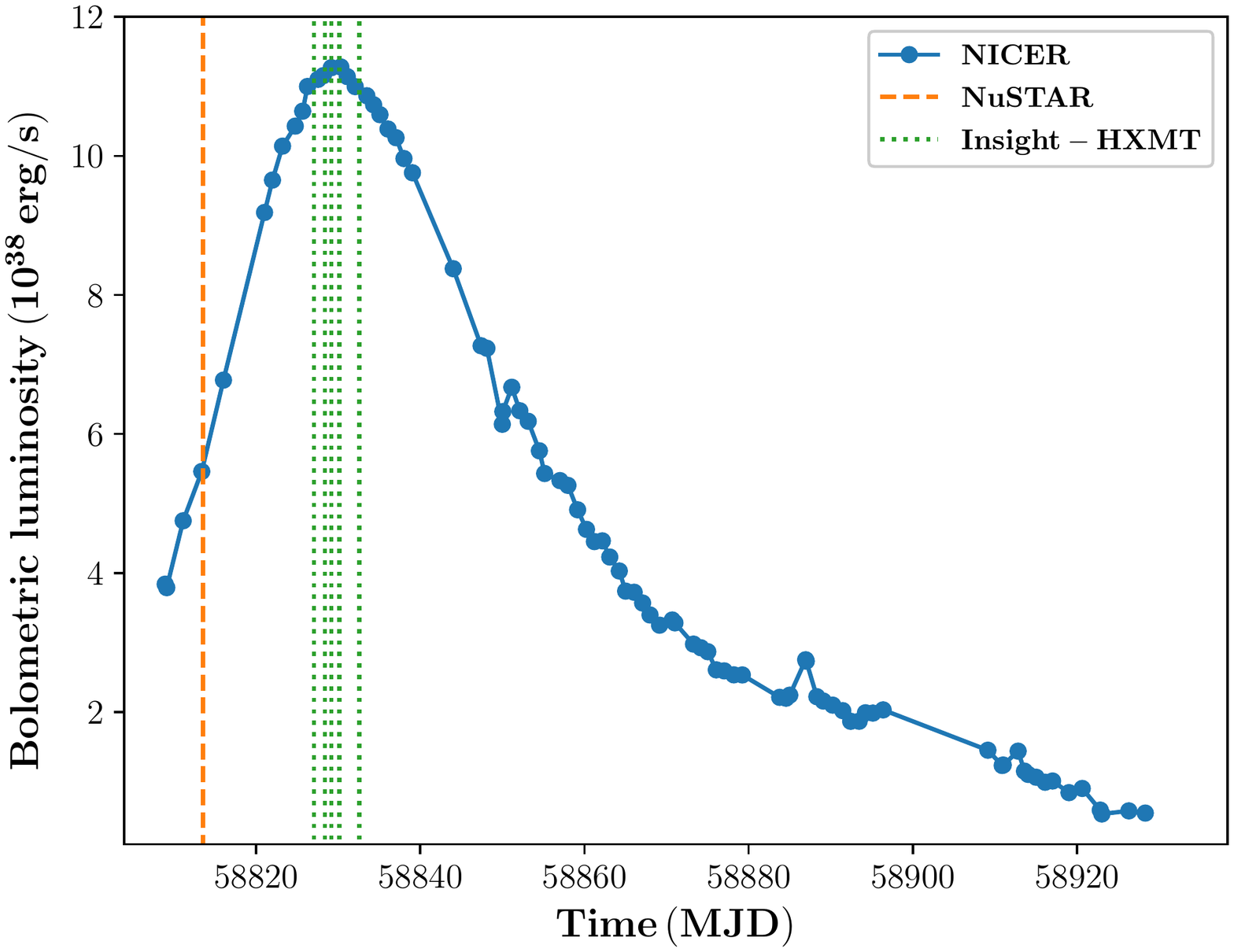}
	\caption{\footnotesize The long-term luminosity evolution of RX J0209.6$-$7427 during its outburst in 2019. The orange (dashed) and green (dash-dotted) lines represent the times of \textit{NuSTAR} and \textit{Insight}-HXMT observations, respectively. }
	\label{fig:nicerlc}
\end{figure}

There are 89 \textit{NICER} observations with the exposure of $>$ 100\,s during the outburst of RX J0209.6$-$7427 in 2019. 
We fit the spectra with a phenomenological model \texttt{Tbabs*(bb+powerlaw)} that describes the data well, and calculate the flux ($F$) in the energy range of $0.5-10$\,keV.
We note that our aim here is to estimate the flux only, and therefore the selection of other alternative models has little influence on our results.
Then we translate $F$ to the ``bolometric" X-ray luminosity (0.5$-$70\,keV) assuming an isotropic radiation as $L=4\,\pi\,C_{\rm bol} F\,D^2 $, where $D$=55\,kpc is the distance to the source \citep{Harries2003,Cignoni2009} and $C_{\rm bol}$ is a conversion factor.
The $C_{\rm bol}$ at the outburst peak is determined using the broadband \textit{Insight}-HXMT observations, and is taken from the previous {\it NuSTAR} report \citep{Vasilopoulos2020} when the source is fainter.
In practice, considering the variation of the spectral shape with luminosity, we estimate this factor using linear interpolation as 
\[C_{\rm bol} = \left\{ \begin{array}{l}
C_{\rm N}, {\rm when}\ F<F_{{\rm N}}\\
C_{\rm N} + (F-F_{\rm N}) \frac{C_{\rm H}-C_{\rm N}}{F_{\rm H}-F_{\rm N}}, {\rm when}\ F\geq F_{{\rm N}}
\end{array} \right.\]
where $F_{\rm H}$ and $F_{\rm N}$ represent fluxes observed with \textit{Insight}-HXMT and {\it NuSTAR} in the energy range of 0.5$-$10\,keV, and $C_{\rm H}$ and $C_{\rm N}$ are the corresponding conversion factors, respectively.
We show the long-term evolution of the bolometric luminosity in Figure~\ref{fig:nicerlc}.
We find the bolometric luminosity is $\sim$ $\rm 10^{39} \, erg\,s^{-1}$ at the outburst peak, making the source a PULX as previously reported.

\begin{figure}[ht!]
	\centering
	\includegraphics[width=0.5\linewidth]{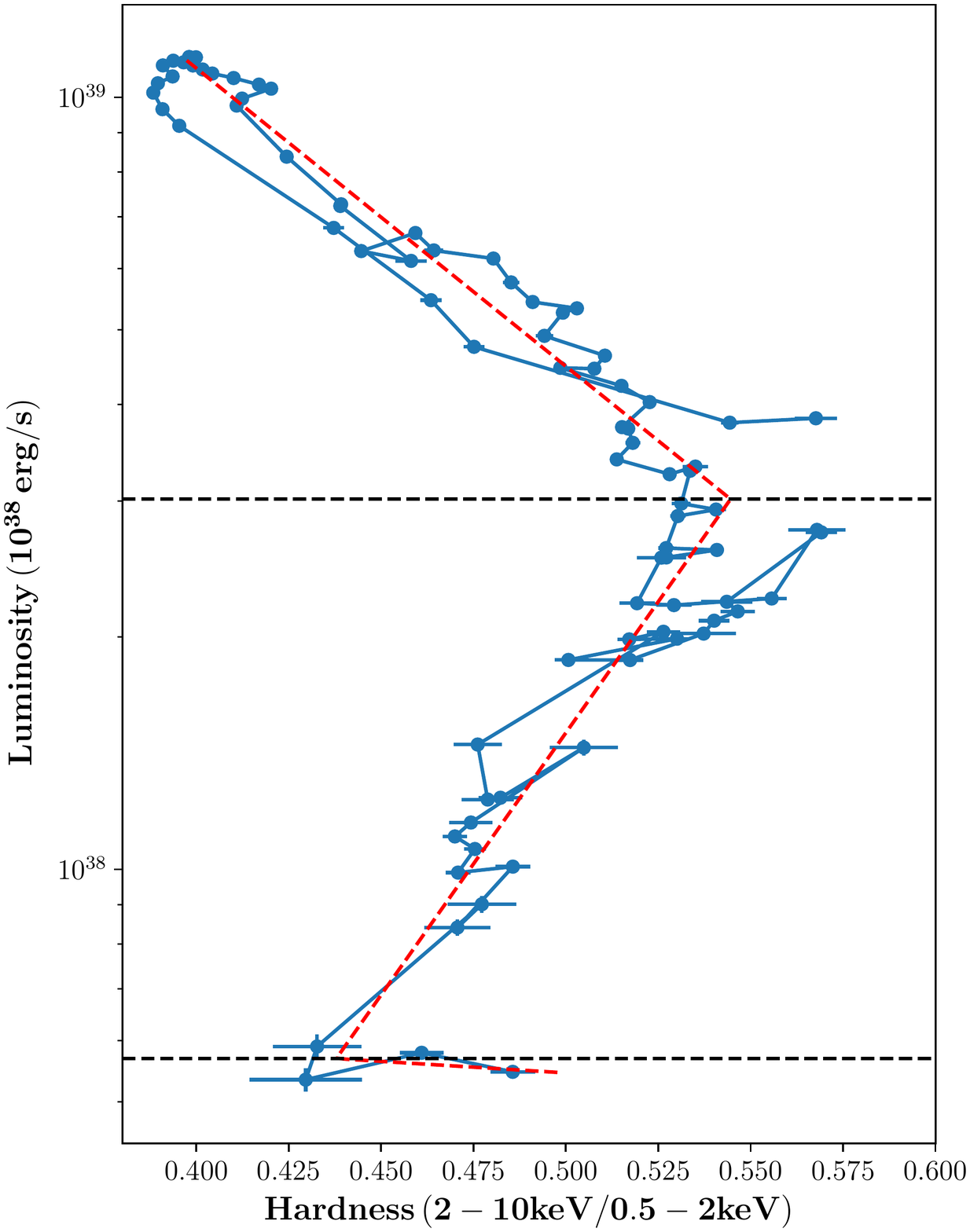}
	\caption{\footnotesize The hardness-intensity diagram of RX J0209.6$-$7427, where the hardness is defined as the count rate ratio of $2-10$\,keV to $0.5-2$\,keV observed with \textit{NICER}. 
	We fitted the HID with broken red lines, which results in two turning points.
	The two dashed black lines represent luminosities of the turning points, i.e., (0.57 $\pm$ 0.01)$\rm \times 10^{38}\,erg\ s^{-1}$ and (3.02 $\pm$ 0.02)$\rm \times 10^{38}\,erg\ s^{-1}$. 
	All the error bars are given at the 68\% confidence level.}
	\label{fig:hardness}
\end{figure}

The Hardness-Intensity Diagram (HID) presents the luminosity and hardness ratio relation of RX J0209.6$-$7427 (Figure~\ref{fig:hardness}), where the hardness is defined as the count rate ratio of 2$-$10\,keV to 0.5$-$2\,keV observed with \textit{NICER}. 
We fitted the HID with broken lines, which results in two turning points at (3.02$\pm$0.02) and (0.57$\pm$0.01) $\rm \times 10^{38}\,erg\ s^{-1}$, respectively. 
Implication of the luminosity turning will be discussed in Section \ref{statetrant}.

\subsection{Timing analysis}
\subsubsection{Pulsation search}
We use the partially phase-coherent analysis to search for pulsations in \textit{NICER} data and generate ephemeris describing the spin and binary parameters of the source \citep{2015ApJ...812...95F}. In practice, we estimated the frequency by folding events on a trial period to obtain an averaged pulse profile for each observation. We then folded 200\,s segments on the same period and cross-correlated the resulting pulse profile with the averaged one to get the time-of-arrival (TOA) of pulsations for each segment. The frequency $\nu$ was then calculated based on these TOAs using the software {\rm {\texttt{Tempo2}} \citep{Hobbs2006}}. The frequency derivative $\nu_{1}$ is estimated from two adjacent observations of \textit{NICER} with the same method.

\begin{figure}[ht!]
	\centering
	\includegraphics[width=0.5\linewidth]{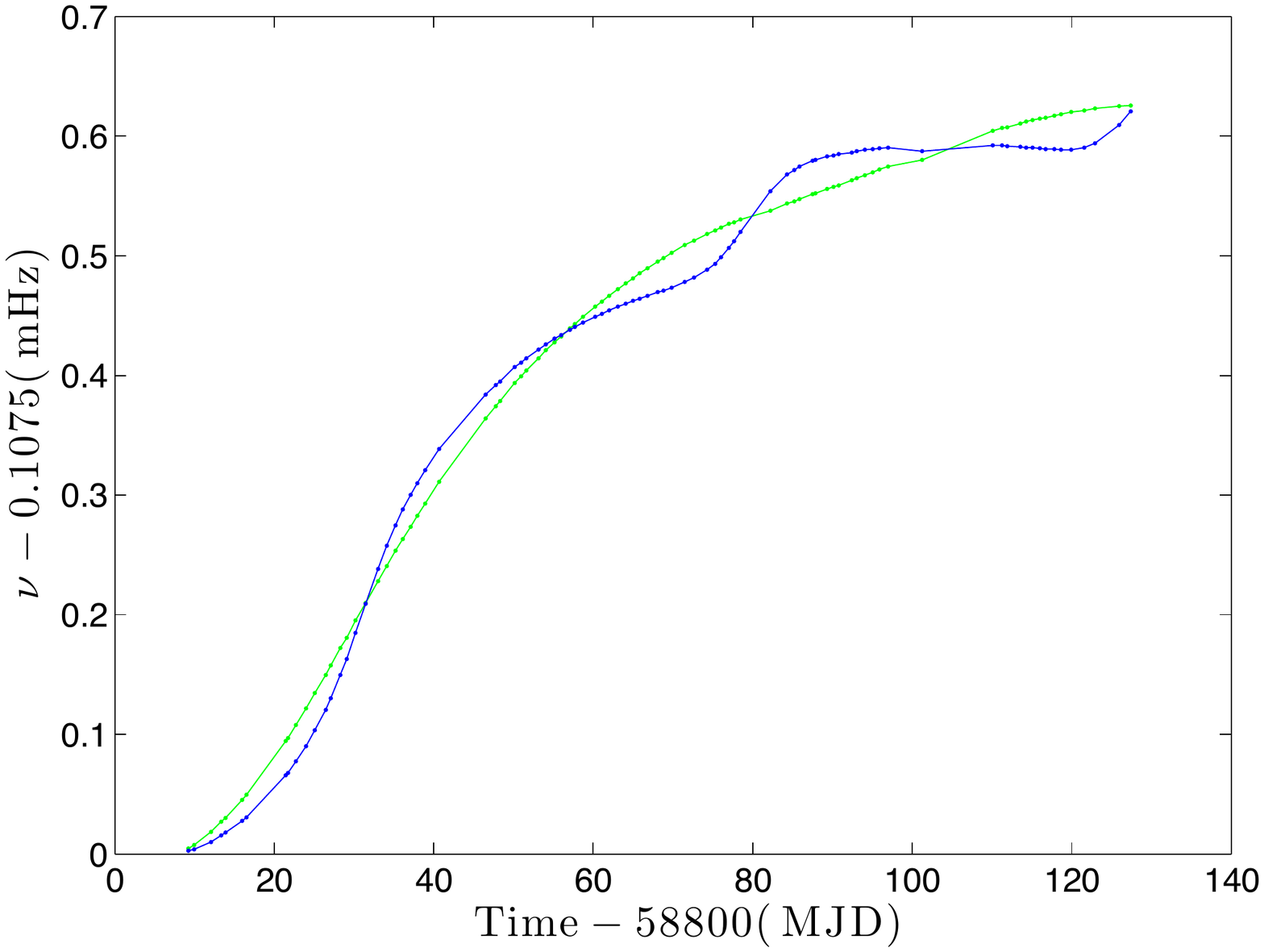}
	\caption{\footnotesize The spin evolution of RX 0209.6$-$7427 with time using \textit{NICER} data. The blue and green points represent the observed spin frequency and that with orbital correction, respectively.}
	\label{fig:spin_evol}
\end{figure}

\begin{figure}[ht!]
	\centering
	\includegraphics[scale=0.5]{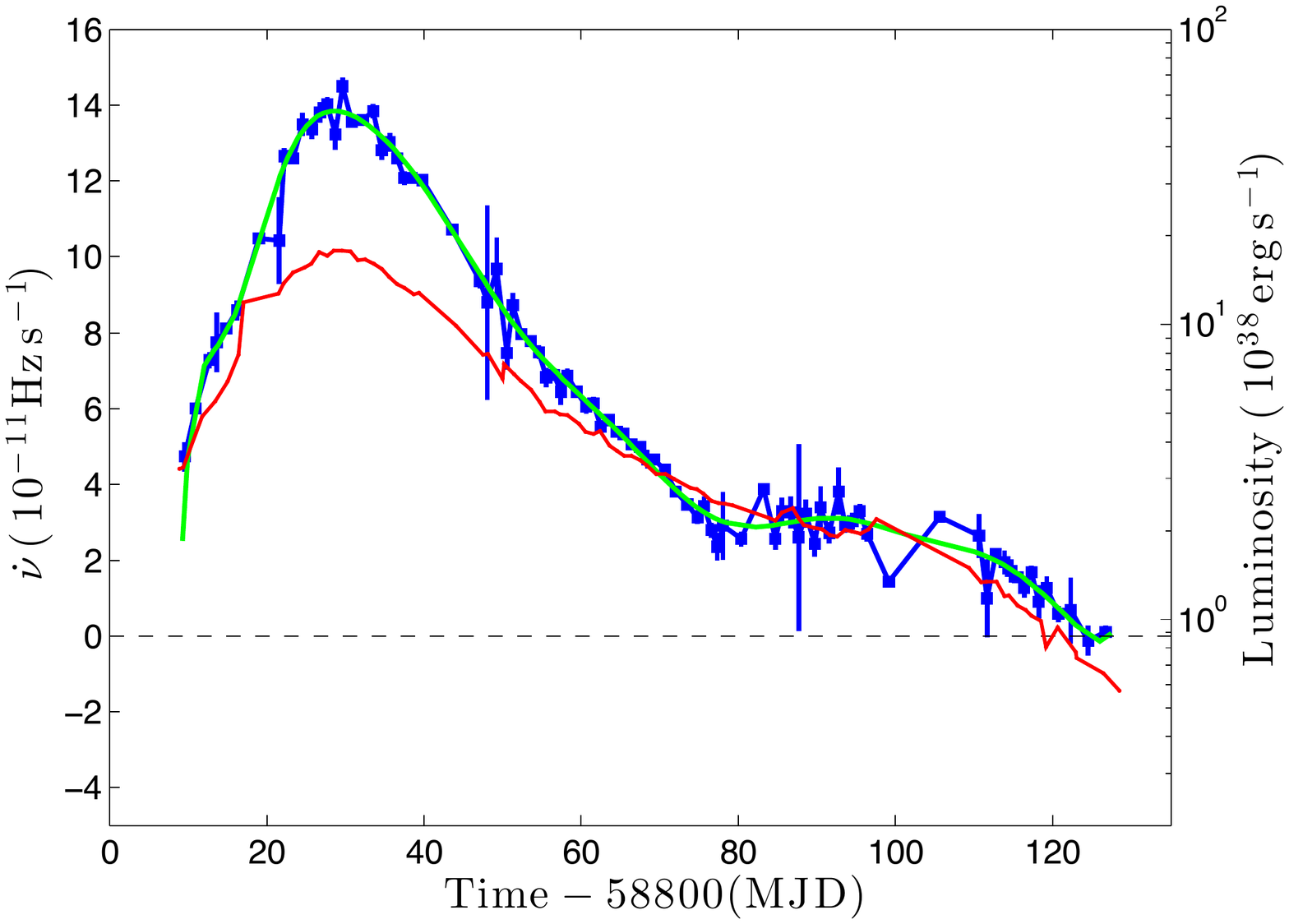}
	\caption{\footnotesize The spin frequency derivative $\dot{\nu}$ as function of time using \textit{NICER} data. The $\dot{\nu}$ values are shown as blue squares. The green solid line is the smoothed line of $\dot{\nu}$ and the red solid line is the bolometric luminosity of the source. The horizontal dashed line represents $\dot{\nu}=0$. All the error bars are given at the 68\% confidence level.}
	\label{fig:rate_evol}
\end{figure}

During the 2019 outburst of RX J0209.6$-$7427, a spin-up trend is significantly found, which is probably caused by the accretion process (Figure~\ref{fig:spin_evol}). On the other hand, there is a frequency variation that shows approximately sinusoidal modulations superposed on the spin-up evolution, which is likely due to the Doppler effect of the binary motion. Thus, the frequency can be described as
\begin{eqnarray}
\nu(t) = \nu(t)_{\rm in} - \nu(t)_{\rm Do}    \, ,    \\
\nu(t)_{\rm in}=\nu_0 + \sum_{n=1}^{6}\frac{1}{n!}\nu_{n}(t-t_0)^n    \, ,    \\
\nu(t)_{\rm Do} = \frac{2\pi\nu_0 a\,{\rm sin}\,i}{P_{\rm orb}} ({\rm cos}\,l + e\,{\rm sin}\,\omega\,{\rm sin}\,2l + e\,{\rm cos}\,\omega\,{\rm cos}\,2l)   \, ,  
\end{eqnarray}
where $\nu(t)_{\rm in}$ and $\nu(t)_{\rm Do}$ represent the intrinsic frequency of the source and the frequency shift caused by the Doppler effect, respectively. 
$\nu_0$ is the frequency at the reference time $t_0$, $a\,{\rm sin}\,i$ is the projected orbital semi-major axis in units of light-travel time, $P_{\rm orb}$ is the orbital period, $e$ is the eccentricity, $\omega$ is the longitude of periastron, and $l$ is the mean longitude. The higher order derivatives $\nu_{n}$ are utilized to fit the spin evolution in order to fit the binary parameters.
We obtained an acceptable fit ($\chi^2$=0.992 with 76 $dof$) using the model mentioned above, and show the results in Figure~\ref{fig:rate_evol} and Table~\ref{table:lcfit}. 

We search independently the spin periods for the five \textit{Insight}-HXMT observations. Due to the short time of \textit{Insight}-HXMT in each observation, spin period derivatives and binary parameters have been ignored in the search, which have in fact negligible effect on the result. We used the cross-correlation technique (See Appendix \ref{highestE}) to determine the highest energy pulsation.

\subsubsection{Pulse profiles}
\begin{figure}[ht!]
	\centering
	\includegraphics[scale=0.45]{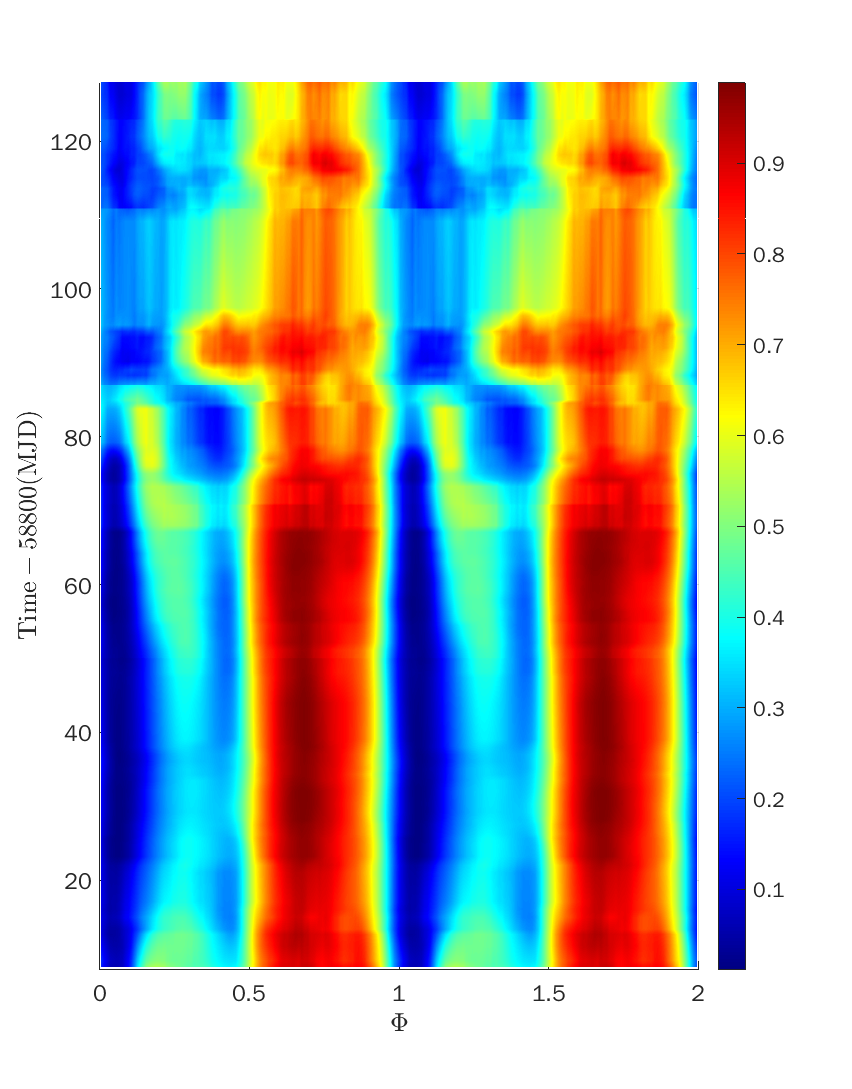}
	\includegraphics[scale=0.45]{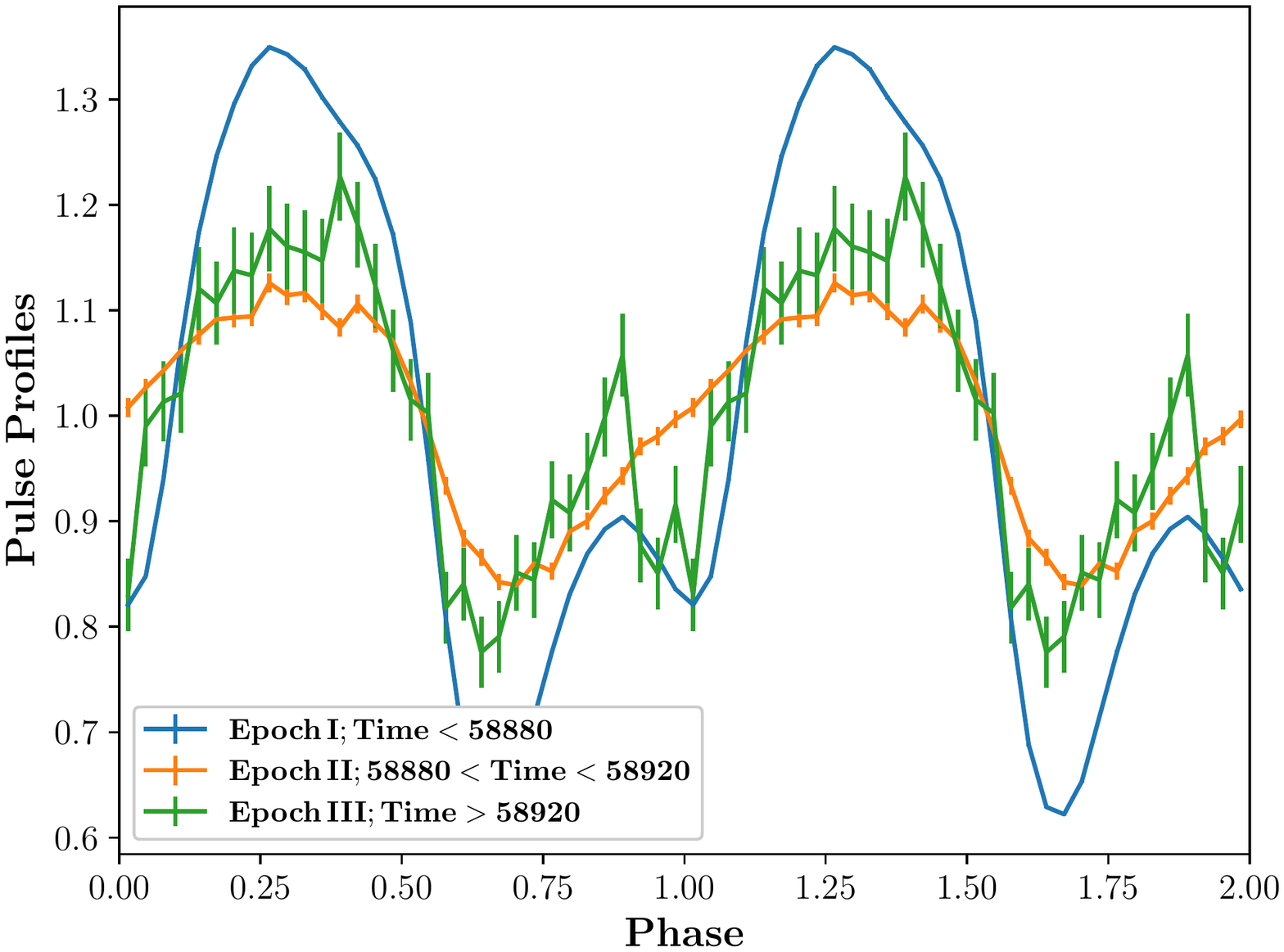}
	\caption{\footnotesize Left: The pulse profile evolution of RX 0209.6$-$7427 with time using \textit{NICER} data. Right: Representative pulse profiles at different epochs using \textit{NICER} data.}
	\label{fig:nicerprofiles}
\end{figure}

We folded events using the resulted timing model as described above to obtain the pulse profile for each \textit{NICER} observation in the energy range of 0.5$-$10\,keV (Figure~\ref{fig:nicerprofiles}, left panel). We found that the morphology of the pulse profile is related to time or the flux: Epoch I, when time $<$ 58880\,MJD, i.e., around the outburst peak, the pulse profile shows one main peak and one minor peak; Epoch II, when 58880\,MJD $<$ time $<$ 58920\,MJD, there is only one broad peak found; Epoch III, when time $>$ 58920\,MJD, i.e., in the faint state of the source, the minor peak appears again. We show averaged pulse profiles using \textit{NICER} data during these three epochs in the right panel of Figure~\ref{fig:nicerprofiles}. 

\begin{figure}[ht!]
	\centering
	\includegraphics[scale=0.53]{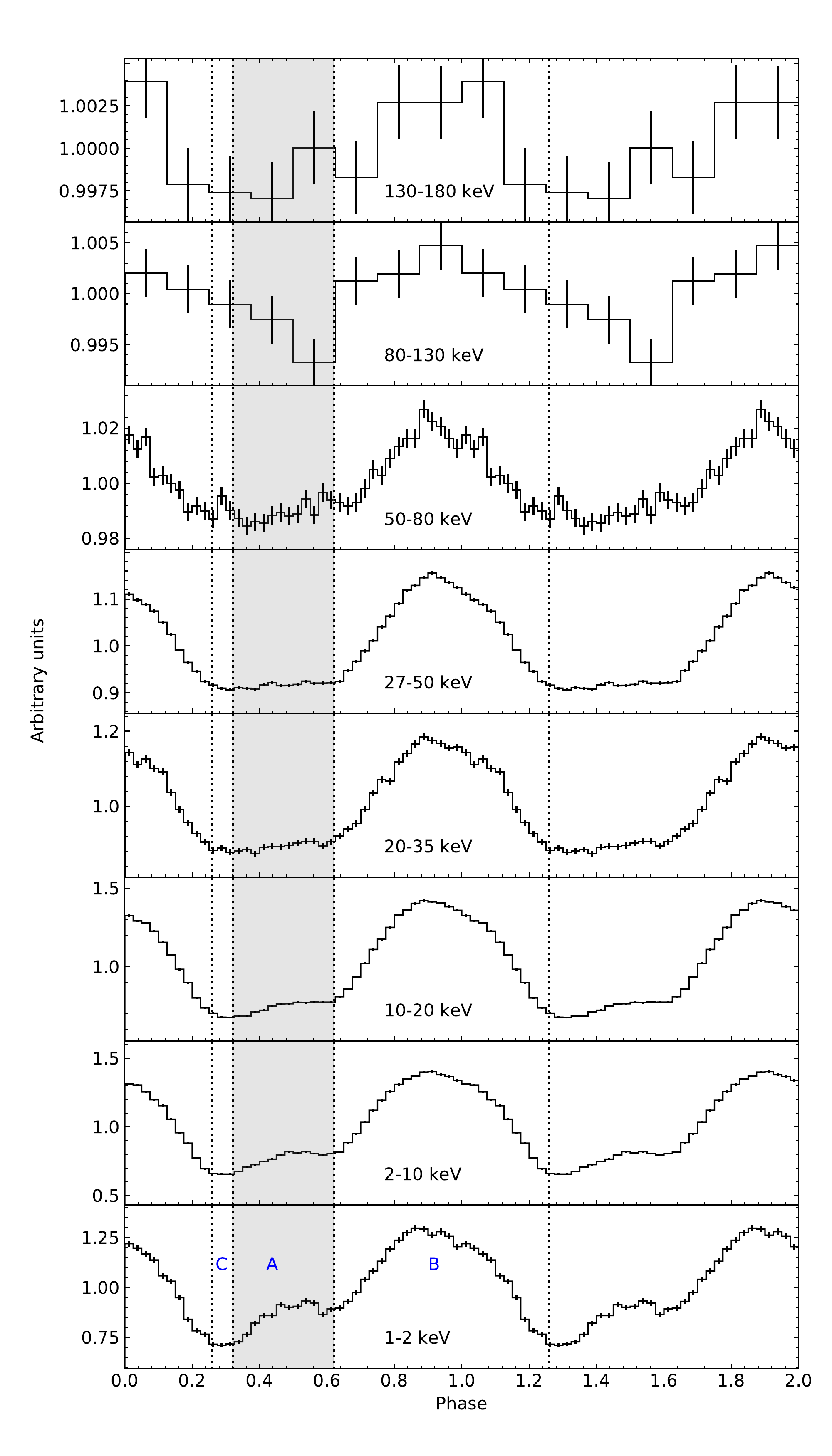}
	\caption{\footnotesize Pulse profiles of RX J0209.6$-$7427 from \textit{Insight}-HXMT observations. The profiles are normalized to the mean value in each energy band and two rotational periods are shown for clarity. Vertical 1$\sigma$ error bars are derived from the counting statistics by assuming Poisson fluctuations. The number of bins in each energy band was simply determined based on the photon statistics. Vertical dotted lines indicate the ON pulse and OFF pulse definitions for both the main and minor peaks. A (0.32-0.62): ON pulse of the minor peak (shaded area); B (0.62-1.26): ON pulse of the main peak; C (0.26-0.32): OFF pulse of the main and minor peaks.} 
	\label{fig:hxmtprofiles}
\end{figure}

The pulsation is detected in all three instruments of \textit{Insight}-HXMT, covering a combined energy range of 1$-$250 keV, as shown in Figure~\ref{fig:hxmtprofiles}. Before 58880 MJD, the pulse profile of RX J0209.6$-$7427 exhibits one main peak and one minor peak separated by $\sim$0.5 in phase. Its main peak is detected above 130 keV and up to 180 keV at a 4.3$\sigma$ confidence level using the cross-correlation technique, the highest energy pulsation detected from this PULX. However, the minor peak becomes weaker with increasing energy and is eventually not prominent above around 27 keV. In comparison, the standard $\chi^2$-test method gave only 1.7$\sigma$ for the 130$-$180 keV profile; this demonstrates that the cross-correlation method is more advantageous since it makes use the lower energy pulse profile, which has very high signal-to-noise ratio, as a template in searching for the higher energy pulsed signals.

To quantify the shape of pulse profiles, we computed the pulse fraction (PF)\footnote{We verified that the root-mean-squared PF formula gave consistent result.} in the energy range of 0.5$-$10\,keV for the \textit{NICER} profile and 1$-$180\,keV for the \textit{Insight}-HXMT profile, respectively: 
\begin{equation}
     PF=(F_{\rm max}-F_{\rm min})/(F_{\rm max}+F_{\rm min})    \, ,
\end{equation}
where $F_{\rm max}$ and $F_{\rm min}$ are the maximum and minimum fluxes in the pulse profile, respectively. We show the result in Figure~\ref{fig:profiles_PF}. The errors were estimated using Monte-Carlo simulations. 
It is clear that the \textit{NICER} PF has a maximum point around 6 $\rm \times 10^{38}\,erg\ s^{-1}$. In addition, we found the slope of the PF evolution changes around the critical luminosity inferred from the hardness variation (Figure~\ref{fig:hardness}). 
Implications of the PF turning will be discussed in Section \ref{statetrant}.

\subsection{Insight-HXMT spectral analysis}
We fitted the 5-observation combined phase-averaged spectra using different phenomenological models available at \texttt{XSPEC} and using the standard \textit{Insight}-HXMT background model. The spectra were best fitted with the \texttt{Tbabs*(cutoffpl+bbodyrad+Gaussian)} model. The \texttt{Tbabs} model is used to account for the photoelectric absorption by the interstellar medium, and the hydrogen column density has been fixed to the Galactic value of $1.58\times 10^{21} \, \rm cm^{-2}$ in the direction of the SMC as was done in previous studies\footnote{We verified that letting the column density free has negligible effect on our result.} \citep{Vasilopoulos2020}. The continuum is composed of a power law with exponential cutoff and a blackbody component. The emission line at $\sim$ 6.4 keV that originates from neutral Fe is modeled by a Gaussian function, but the limit of the \textit{Insight}-HXMT spectral resolution does not allow more detailed modeling of the emission line to probe its physical origin. The best fitting parameters are presented in Table~\ref{table:fitave} and the spectra are shown in Figure~\ref{fig:avespec_hxmtbkg}. The broad band (1$-$150 keV) luminosity is $(1.11\pm 0.06)\times 10^{39}\, \rm erg\ s^{-1}$.

The phase-resolved spectra using the definition of ON (pulsed) and OFF (unpulsed) ranges in the pulse profile for both the main and minor peaks (Figure~\ref{fig:hxmtprofiles}) were created using standard \textit{Insight}-HXMT data analysis procedure and corrected for exposure in different phases. Both the standard \textit{Insight}-HXMT background model \citep{Liao2020} and the OFF spectrum have been used to estimate the background emission. We tried different phenomenological models available in \texttt{XSPEC} to fit the main and minor peaks. When using the standard \textit{Insight}-HXMT background model, the best-fit models for the main and minor peaks are \texttt{Tbabs*(cutoffpl+bbodyrad+bbodyrad+Gaussian)} and \texttt{Tbabs*(cutoffpl+bbodyrad+Gaussian)}, respectively. 

When using the OFF spectrum as background, the best-fit models for the main and minor peaks are \texttt{Tbabs*(cutoffpl+bbodyrad)} and \texttt{Tbabs*cutoffpl}, respectively. The blackbody component for the main peak, about 1.4\% of the total flux, has a temperature of $kT \sim 2$\,keV and a size of $\sim 3$\,km, possibly accounting for the thermal emission from a hot spot on the NS surface. When using the \textit{Insight}-HXMT background model, an additional thermal component ($kT_{\rm bb} \sim 0.2$\,keV, $R_{\rm BB} \lesssim 10^3$\,km), about 0.13\% of the total flux, is required for the spectra of both the main and minor peaks. By considering a magnetospheric radius $R_{\rm m} \sim$1000 km with a dipole magnetic field of 4.8$\times10^{12}$ G (see Section \ref{statetrant} for details), this thermal component is likely from the inner part of the accretion disk truncated by the magnetosphere. The inner disk is illuminated and heated up by the fan beam emission of the accretion column (AC), and then can emit thermal X-ray photons. Moreover, since the disk component appears equally in each phase, it should be eliminated when using the OFF spectrum as background, consistent with our spectral fitting results. The best-fit parameters are presented in Tables~\ref{table:fitbigpeak}-\ref{table:fitsmallpeak} and the spectra are shown in Figure~\ref{fig:bigspec}-\ref{fig:smallspec}.


\section{Discussion} \label{discuss}

\subsection{Emission pattern}
\label{pattern}
In accreting X-ray pulsars, the accretion disk is truncated at the magnetospheric radius ($R_{\rm m}$) where the ram pressure of the matter in the disk is balanced by the magnetic pressure. After that, the matter is guided by the magnetic field lines and funnelled on to polar caps of the NS to generate X-rays. It is believed that the radiation structure near the surface of the NS depends on luminosity and accretion rate \citep{Basko1975, Basko1976,Becker2007, Becker2012, Mushtukov2015b}. When the source is faint (subcritical accretion), most of the radiation escapes along the magnetic field lines forming a ``pencil beam" pattern \citep{Basko1975,Burnard1991,Nelson1993,Becker2012}. When the source is brighter, an optically thick AC begins to rise above the polar cap, and the radiation escapes mostly from the wall of the AC, i.e., perpendicular to the magnetic field, forming a ``fan'' beam pattern \citep{Davidson1973,Basko1976,Becker2012,Mushtukov2015b}. The separation of the two accretion regimes is defined as the so-called ``critical luminosity" ($L_{\rm crit}$). 

However, the observed emission properties like pulse profiles, spectral shapes and cyclotron resonance scattering feature (CRSF) depend on luminosity, energy, the emission geometry, etc., thus implying a much more complex emission pattern than this simple pencil/fan beam scenario. As we have demonstrated (see Appendix \ref{ACmodel}), lower energy photons can contribute to both beam patterns, while higher energy photons will preferentially escape in the form of fan beam. Therefore, profiles at higher energies are more useful to discriminate the emission patterns.

Applying the commonly used model of AC to RX J0209.6$-$7427, we can demonstrate that the $\sim$130 keV pulsation during the peak of its outburst can only originate from the fan beam of the AC (see Appendix \ref{ACmodel}), which does not have significant effect of beaming toward certain direction. This indicates that the observed luminosity of RX J0209.6$-$7427 is intrinsic. Therefore, the value of $L_{\rm crit}$ we identified from the profile, hardness and PF evolution is robust (see Section \ref{statetrant}), which in turn can result in a robust magnetic field estimation (see Section \ref{mag_estimate}). 

In addition, we identified a rapid decline of PF using \textit{Insight}-HXMT data with increasing energy above 50 keV (Figure~\ref{fig:profiles_PF}). Similar trend of PF decrease has been reported in the \textit{AstroSat} analysis of RX J0209.6$-$7427 as well \citep{Chandra2020}. We propose that this is mostly likely due to the increased fraction of reflection \citep{Poutanen2013} of higher energy photons off the NS surface.  The reflected photons will smear/broaden the pulse peak and thus reduce the PF at high energies. A recent Monte-Carlo simulation shows that the reflected flux off a NS surface increases rapidly with energy and reaches a maximum just before the CRSF \citep{Kylafis2021} (see Fig. 2 in their paper, where the cyclotron energy was taken as around 25 keV). Therefore the reflection fraction increases with energy (below CRSF) and then at some points, it becomes dominant to reduce the observed PF. For RX J0209.6$-$7427, the CRSF energy is probably at around 200 keV or above (see Section~\ref{mag_estimate} for magnetic field estimates), though a CRSF has not been detected yet, most likely due to the limited sensitivity of the current X-ray telescopes above 100 keV. Therefore the PF decrease with energy above about 50 keV is consistent with the increased reflection fraction with increasing energy before reaching the CRSF energy. 

Actually, the simulations in \cite{Kylafis2021} did not consider the energy dependence of the emission geometry of the AC, as shown in Figure~\ref{fig:geometry}, where the emission region of higher energy photons in the form of fan beam is closer to the surface of the NS and thus can illuminate the NS surface even more easily. The rapid PF decrease with energy above around 50 keV (and below the possible CRSF energy) is likely due to the combination of the reflection physics simulated in \cite{Kylafis2021} and the geometrical effect illustrated in Figure~\ref{fig:geometry}. Further detailed modeling and simulation are required to fully explain the broad-band PF evolution with energy, which may provide further confirmation of the fan-beam nature of the high energy pulsed emission.

\subsection{State transition}
\label{statetrant}
\begin{figure}[ht!]
	\centering
	\includegraphics[scale=0.45]{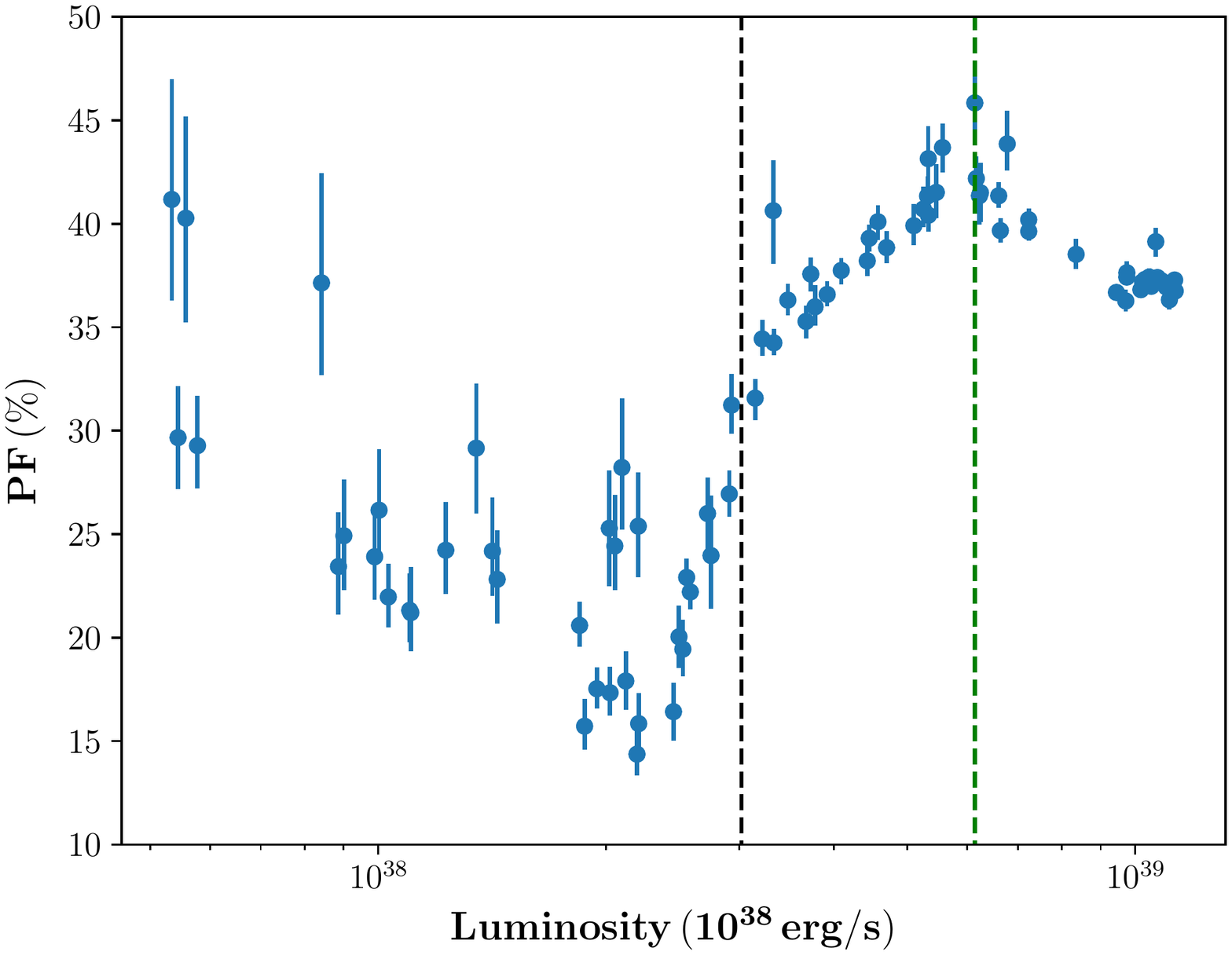}
	\includegraphics[scale=0.5]{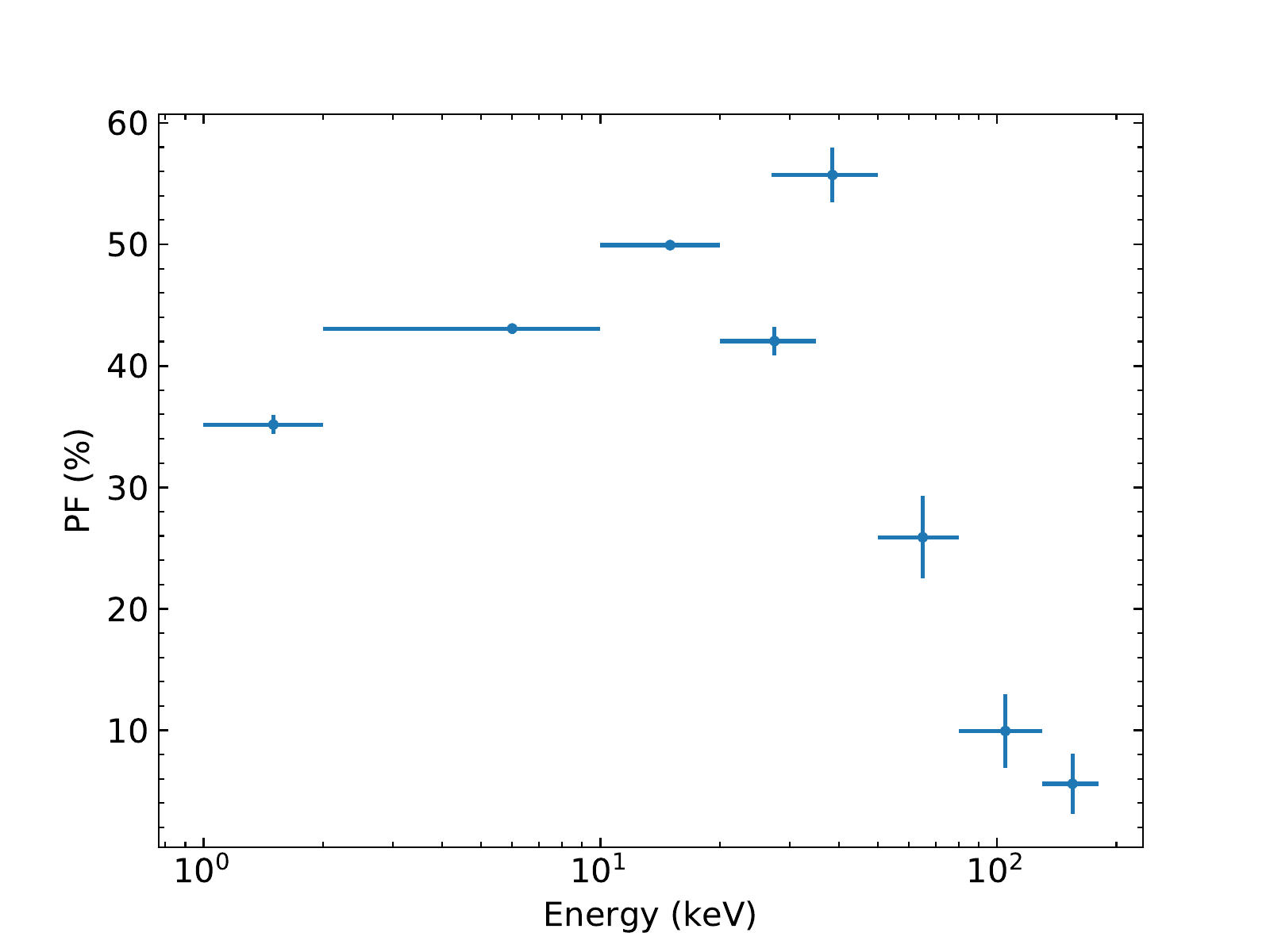}
	\caption{\footnotesize Left: the evolution of the PF using \textit{NICER} data. 
	The vertical dashed green line shows the luminosity where the PF evolution changes from increasing to decreasing.
	The vertical dashed black line presents the critical luminosity suggested by Figure~\ref{fig:hardness}, where the increasing slope of the PF becomes flatter.
	Right: the energy dependence of PF from 1 keV to above 130 keV using \textit{Insight}-HXMT data. All the error bars are given at the 68\% confidence level. Note that the high PF in 27$-$50 keV is most likely due to the prominent peak around 30$-$40 keV in the \textit{Insight}-HXMT background model  (see Figures 5 and 10 in \cite{Liao2020}) which causes an over-subtraction of the background when calculating the PF.}
	\label{fig:profiles_PF}
\end{figure}

\begin{figure}[ht!]
	\centering
	\includegraphics[scale=0.6]{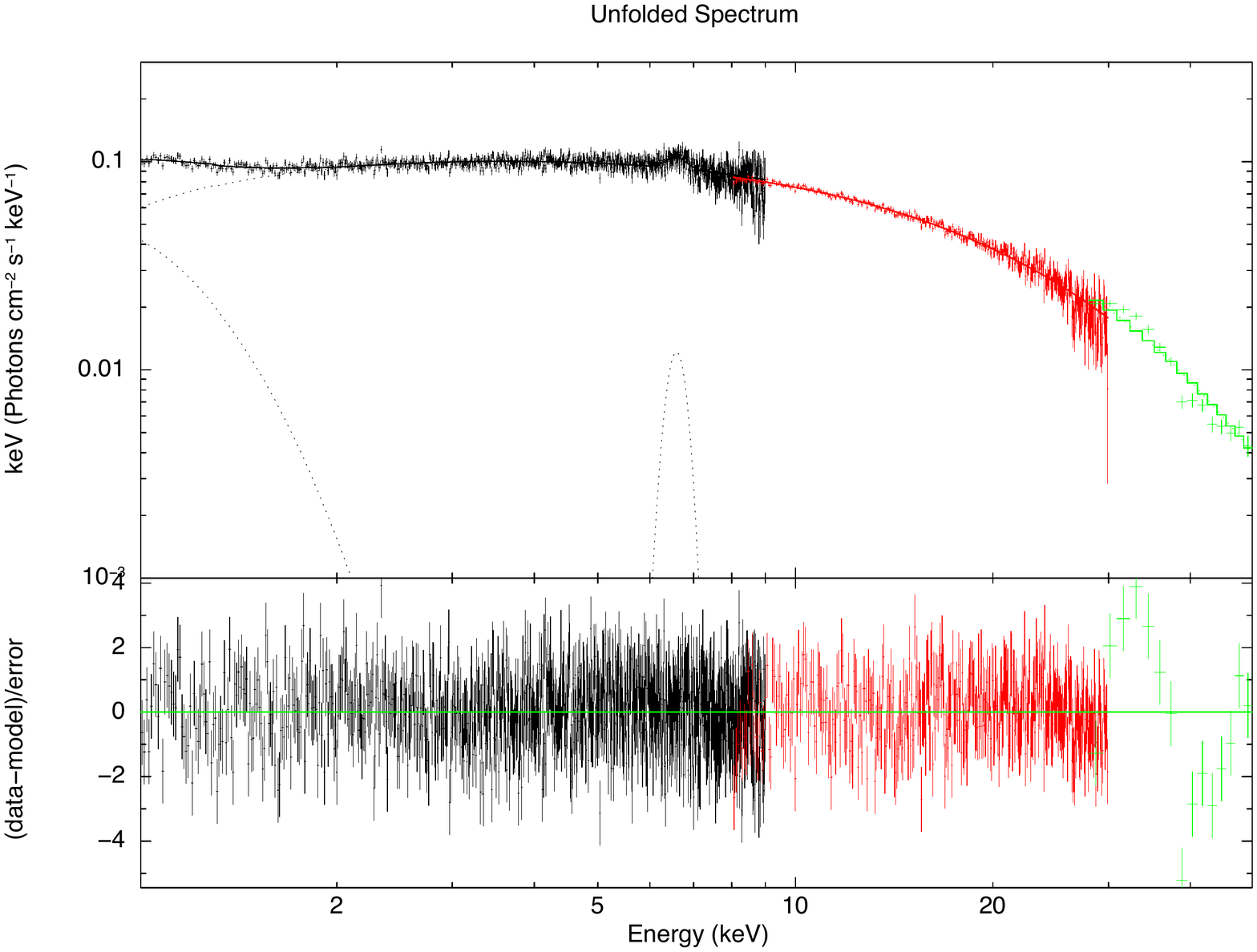}
	\caption{\footnotesize The 5-observation combined phase-averaged spectrum of RX J0209.6$-$7427 observed with the \insight{} during the 2019 outburst in the energy band of $1-50$ keV fitted with the \texttt{Tbabs*(cutoffpl+bbodyrad+Gaussian)} model using the \insight{} background model. Above 50 keV, the \insight{} background uncertainty ($\sim$1\%) is comparable to the flux of the source emission, so it is inappropriate to include data above 50 keV. The absorption-like feature at around 40 keV in the residuals is due to the structure in the background model around this energy. The lower panel shows the residuals to the best-fitting model.}
	\label{fig:avespec_hxmtbkg}
\end{figure}

\begin{figure}[ht!]
	\centering
	\includegraphics[scale=0.4]{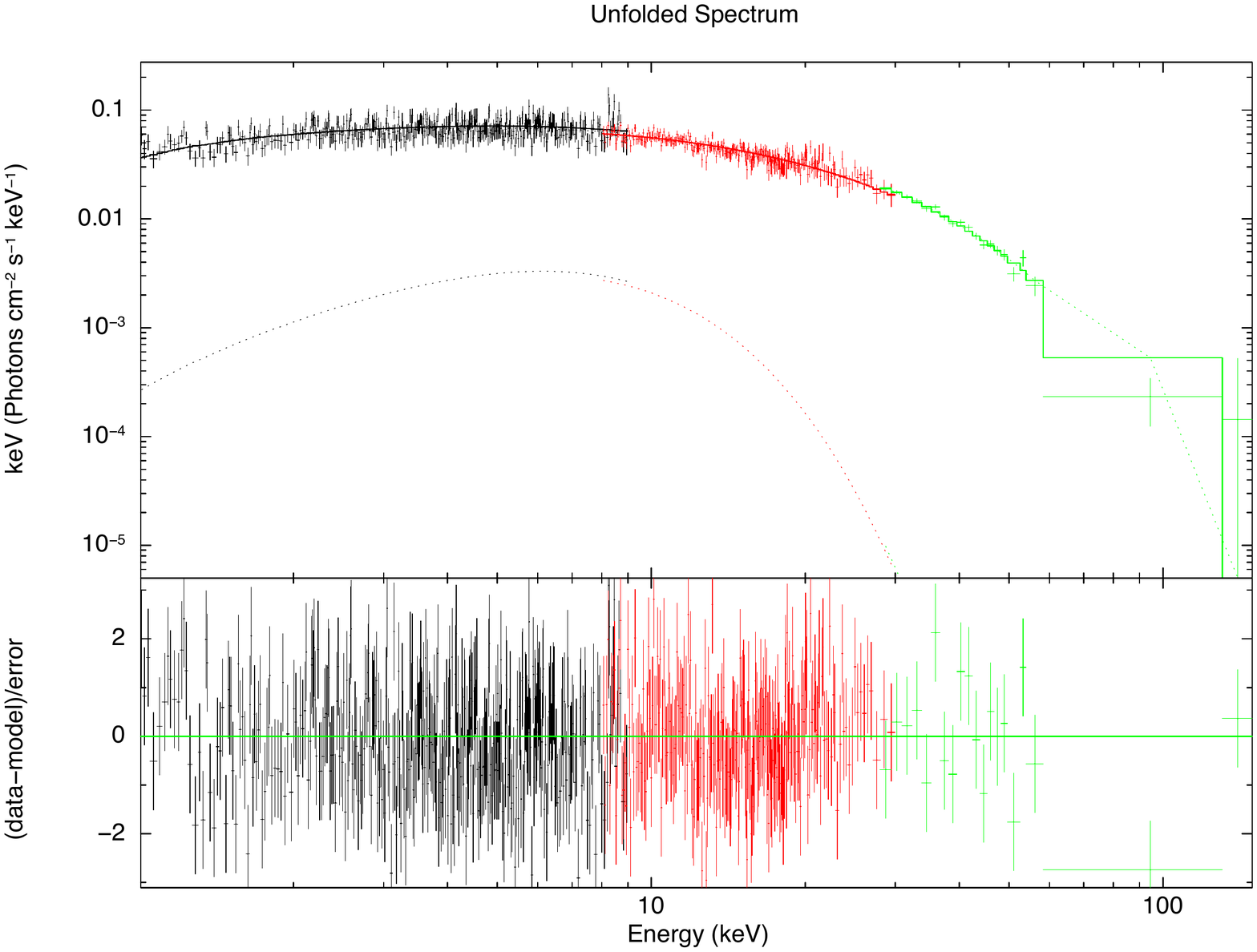}
	\includegraphics[scale=0.4]{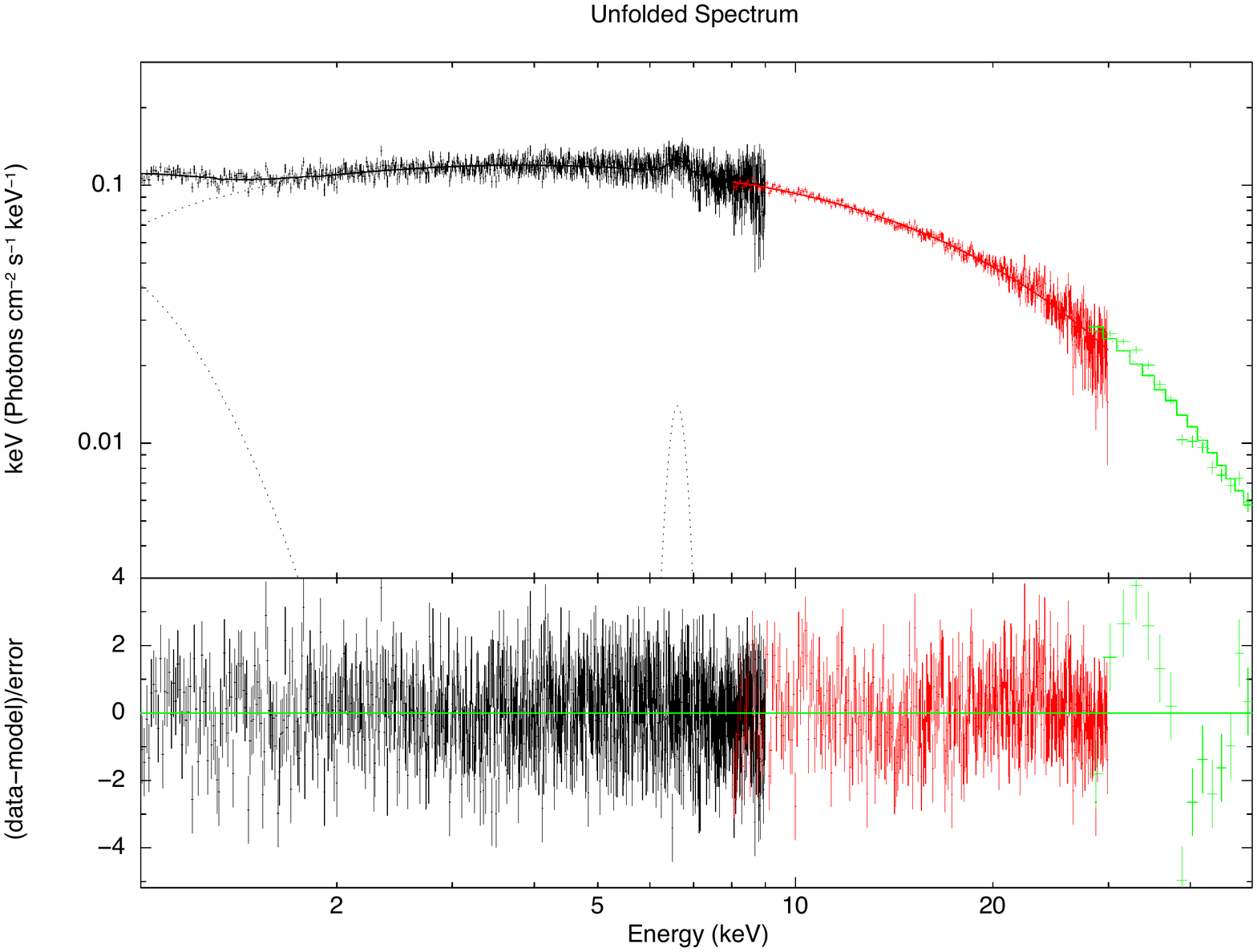}
	\caption{\footnotesize The 5-observation combined spectrum of the main peak of RX J0209.6$-$7427 observed with the \insight{} during the 2019 outburst. The lower panel of each spectrum shows the residuals to the best-fitting model. Top: Fitted using the OFF spectrum as background in the energy band of $1-150$ keV, the best-fit model is \texttt{Tbabs*(cutoffpl+bbodyrad)}. Bottom: Fitted using the \insight{} background model in the energy band of $1-50$ keV. The best-fit model is \texttt{Tbabs*(cutoffpl+bbodyrad+bbodyrad+Gaussian)}. Above 50 keV, the \insight{} background uncertainty ($\sim$1\%) is comparable to the flux of the source emission, so it is inappropriate to include data above 50 keV. The absorption-like feature at around 40 keV in the residuals is due to the structure in the background model around this energy. }
	\label{fig:bigspec}
\end{figure}

\begin{figure}[ht!]
	\centering
	\includegraphics[scale=0.4]{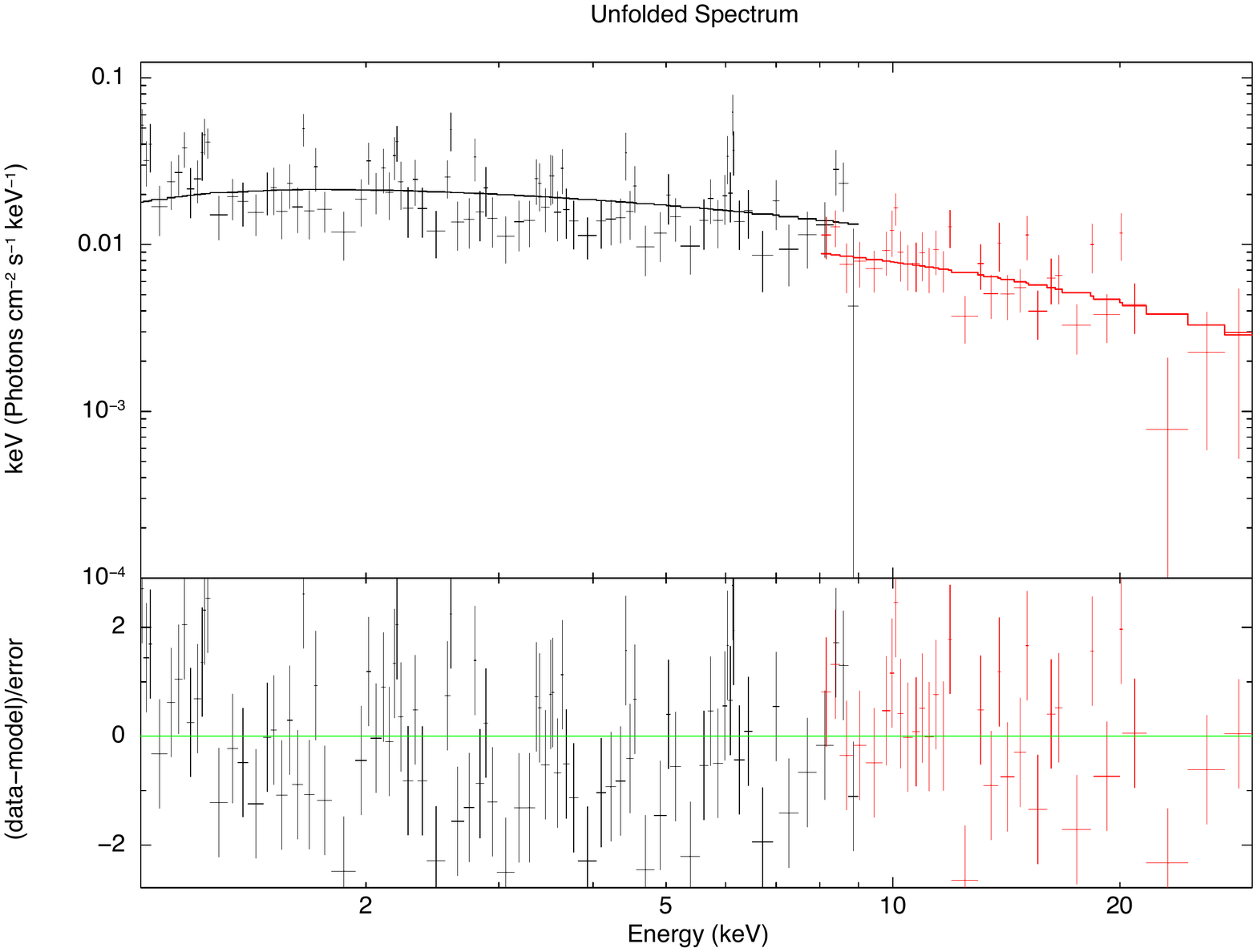}
	\includegraphics[scale=0.4]{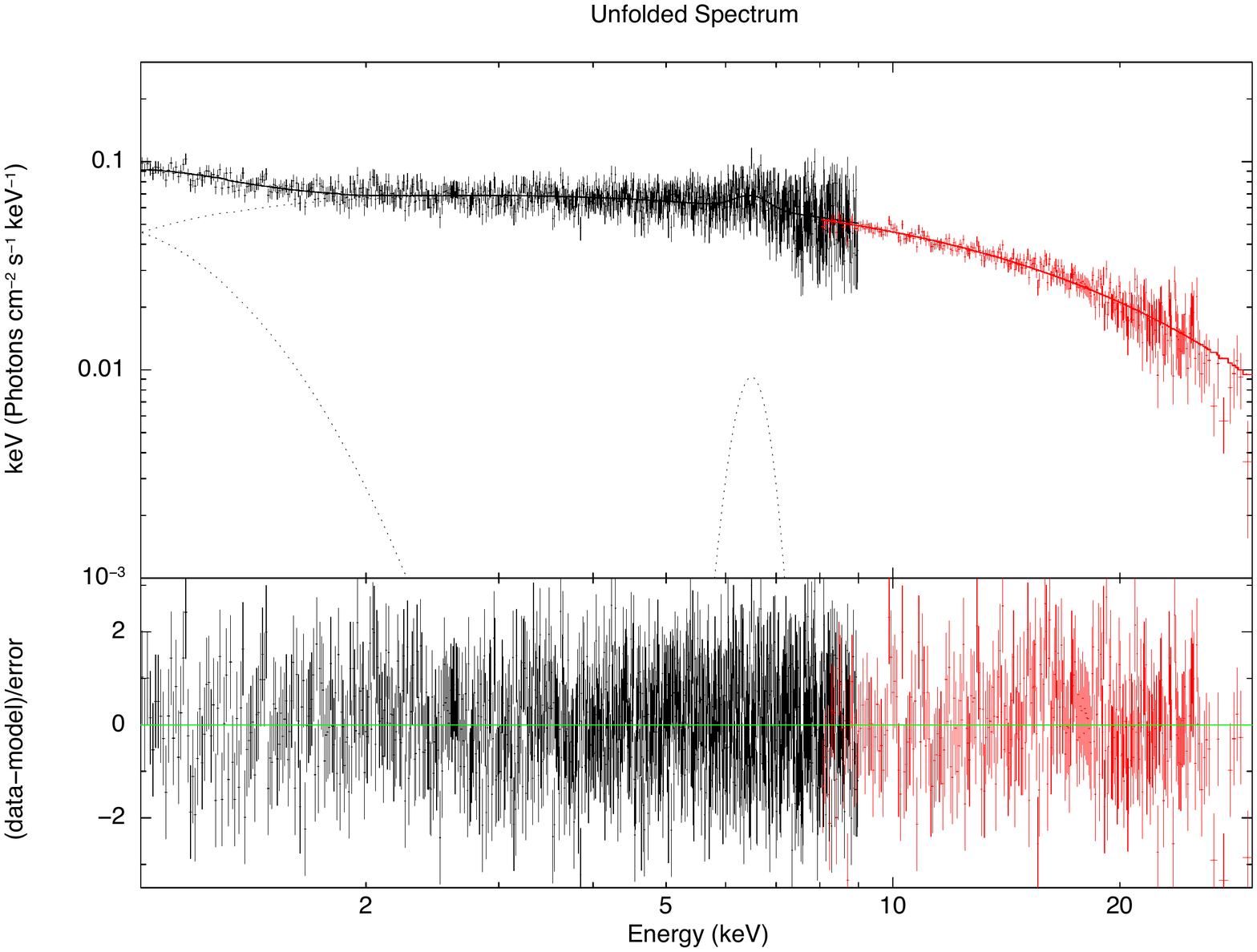}
	\caption{\footnotesize The 5-observation combined spectrum of the minor peak of RX J0209.6$-$7427 observed with the \insight{} during the 2019 outburst in the energy band of $1-30$ keV. The lower panel of each spectrum shows the residuals to the best-fitting model. Top: Fitted using the OFF spectrum as background, the best-fit model is \texttt{Tbabs*cutoffpl}. Bottom: Fitted using the \insight{} background model, the best-fit model is \texttt{Tbabs*(cutoffpl+bbodyrad+Gaussian)}.}
	\label{fig:smallspec}
\end{figure}

With the change of accretion rate, significant transitions of the spectral shape, pulse profile and cyclotron line evolution have been expected around $L_{\rm crit}$ in theory and observed in a large number of sources \citep{Sasaki2012,Reig2013,Postnov2015,Doroshenko2017,Wilson2018,Doroshenko2020,Ji2020,Doroshenko2020,Kong2020}. We have investigated the temporal and spectral properties of RX J0209.6$-$7427 and found significant changes of its hardness
around 3.02$\pm$0.02 and (0.57$\pm$0.01)$\rm \times 10^{38}\,erg\ s^{-1}$ (Figures \ref{fig:hardness}, \ref{fig:nicerprofiles} and \ref{fig:profiles_PF}) during the outburst. Following the arguments in \cite{Reig2013}, the first turning luminosity of (3.02$\pm$0.02) $\times 10^{38}\,{\rm erg\ s}^{-1}$ which is the separation of the two branches in the HID may correspond to the transition from the supercritical to the subcritical regimes with the disappearance of the AC, similar to what was found in the Galactic PULX Swift J0243.6+6124 \citep{Doroshenko2020,Kong2020}. This tuning luminosity is therefore considered to be $L_{\rm crit}$.
In addition, around this luminosity it seems that the evolution of the PF changes as well (Figure~\ref{fig:profiles_PF}), which is consistent with theoretical expectations \citep[e.g.,][]{Basko1976}.
Under this interpretation, it is clear that the source was already in the supercritical state when \textit{NICER} started the series of observations. 
We propose that the second turning luminosity represents the transition from the subcritical regime to the state when the gas shock disappears or becomes non-dominant \citep{Becker2012}, while this was not found in Swift J0243.6+6124.

On the other hand, we found a turning point where the \textit{NICER} PF evolution changes from increasing to decreasing (Figure \ref{fig:profiles_PF}, left panel) around 6 $\rm \times 10^{38}\,erg\ s^{-1}$. We suggest that this may correspond to the transition between the radiation pressure dominated (RPD) disk and the gas pressure dominated (GPD) disk as theoretically proposed \citep{Shakura1973,Mushtukov2015a}, similar to the case of Swift J0243.6+6124 \citep{Doroshenko2020,Kong2020}. Following the methodology presented in \cite{Doroshenko2020}, we can compare the magnetospheric radius \citep{Andersson2005,Mushtukov2015a,Campana2018,Monkkonen2019}
\begin{equation}
    R_{\rm m} = 2.6\times 10^8\, k\, m^{1/7}\,R_{6}^{10/7}\,B_{12}^{4/7}\,L_{37}^{-2/7}  \,\,\, \rm cm  
\end{equation}
with the boundary between the GPD and RPD zones \citep{Monkkonen2019}
\begin{equation}
    R_{\rm AB} = 10^7\, m^{1/3}\,\dot{M}_{17}^{16/21}\,\alpha^{2/21}  \,\,\, \rm cm  \, ,
\end{equation}
where $m$, $R_{6}$, $B_{12}$, $L_{37}$ and $\dot{M}_{17}$ are the mass, radius, magnetic field, luminosity and accretion rate in units of $M_{\rm \odot}$, $10^6$ cm, $10^{12}$ G, $10^{37} \,\rm erg\ s^{-1}$ and $10^{17}\, \rm g\ s^{-1}$, respectively. We assume a typical value of $\alpha=1$. The parameter $k$ is model dependent and is defined as the ratio between $R_{\rm m}$ and the Alfv\'en radius ($R_{\rm A}=(\frac{\mu^4}{GM\dot{M}^2})^{1/7}$). The transitional luminosity from the GPD to the RPD disk is \citep{Andersson2005,Monkkonen2019} 
\begin{equation}
    L_{\rm AB} = 3\times 10^{38}\, k^{21/22}\, \alpha^{-1/11}\, m^{6/11}\, R_{6}^{7/11}\,B_{12}^{6/11} \,\,\,  \rm erg\ s^{-1}  \,\, .
\end{equation}

In this work, we consider that the NS has a radius of $R=10^6\,{\rm cm}$ and a mass of $1.4\,\rm  M_{\odot}$. Using the dipole magnetic field of about 4.8$\times10^{12}$ G estimated from the GL model \citep{Ghosh1979} with $k=0.52$ (Section \ref{mag_estimate}), we got $R_{\rm AB}\sim R_{\rm m} \sim$ 1000 km which is indeed the condition for the transition, and $L_{\rm AB}=5.5\times10^{38} \,\rm erg\ s^{-1}$. This transitional luminosity is consistent with that inferred from the PF evolution.

This interpretation should be, however, considered with caution, given that in Swift J0243.6+6124, both pulse profile and power spectrum changes have been observed to accompany the transition, while in RX J0209.6$-$7427 the power spectrum seems to be consistent with a single power law throughout the outburst and no breaks occurred. In comparison, 2S 1417$-$624 was proposed to transit between GPD and RPD based on the profile changes, but no significant power spectrum change has been detected either \citep{Ji2020b}; while in GRO J1744$-$28, a transition from the GPD disk to the RPD disk was proposed to explain the observed power spectrum change which differs dramatically from the canonical shape for a continuous GPD disk \citep{Monkkonen2019}. The power spectrum change of GRO J1744$-$28 is, however, different from what was found in Swift J0243.6+6124. Such transitions appear to be quite complicated and the origins of which are still not well known. Investigating the properties of the RPD disk in detail is beyond the scope of the paper.

\subsection{Magnetic field estimates}
\label{mag_estimate}
Different approaches have been adopted to estimate the magnetic field of the NS in RX J0209.6$-$7427. (1) Relation between the magnetic field and $L_{\rm crit}$. In theory $L_{\rm crit}$ depends on the magnetic field. It is thus important to measure the value of $L_{\rm crit}$ accurately in order to estimate the magnetic field, which requires an accurate measurement of the distance. This is difficult for sources in the Milky Way but for RX J0209.6$-$7427, its known distance and much smaller relative distance uncertainty are a big advantage. In literature, there are several attempts to compute $L_{\rm crit}$ \citep{Wang1981b, Becker2012, Mushtukov2015b}. For example, \cite{Becker2012} proposed that $L_{\rm crit}\sim 1.49\times 10^{37}B_{12}^{16/15}{\rm erg\,s^{-1}}$. This suggests that the magnetic field of this source is $(1.68\pm0.01$)$\times10^{13}$\,G if $L_{\rm crit}$ is ($3.02\pm0.02$)$\times 10^{38}\,\rm erg\ s^{-1}$ as we determined in Section \ref{lcHID}. On the other hand, according to the model proposed by \cite{Mushtukov2015b} which takes into account different polarization modes, this $L_{\rm crit}$ value requires $E_{{\rm cyc}}>$200 keV, which results in a surface magnetic field strength of higher than 2.2$\times10^{13}$ G \citep[see Figure 5 in][private communication with Alexander Mushtukov]{Mushtukov2015b}. This is similar to the findings in SMC X-3 \citep{Tsygankov2017}.

(2) The magnetic field of the source can be investigated according to the spin-up during the outburst.
Assuming the GL model \citep{Ghosh1979}, the spin-up rate of the NS can be written as:
\begin{equation}
\dot{\nu} = 2^{-15/14}k^{1/2}\mu^{2/7}(GM)^{-3/7}(I\pi)^{-1}R^{6/7}L^{6/7}n(\omega)\ {\rm Hz\,s^{-1}}    \, ,
\end{equation}
where $I=\frac{2}{5}MR^2$, $\mu=\frac{1}{2}BR^3$ and $k$ are the moment of inertia, the magnetic dipole moment, and the dimensionless constant introduced previously. 
We note that $n(\omega)$ is related to the fastness parameter $\omega \equiv (R_{\rm m}/R_{\rm co})^{3/2}$, where $R_{\rm co} \equiv(GMP^2/4\pi^2)^{1/3}$ is the co-rotation radius: 
\begin{equation}
n(\omega) \approx1.39\frac{1-\omega \left[4.03(1-\omega)^{0.173}-0.878\right]}{1-\omega} .
\end{equation}

\begin{figure}[ht!]
	\centering
	\includegraphics[width=0.6\linewidth]{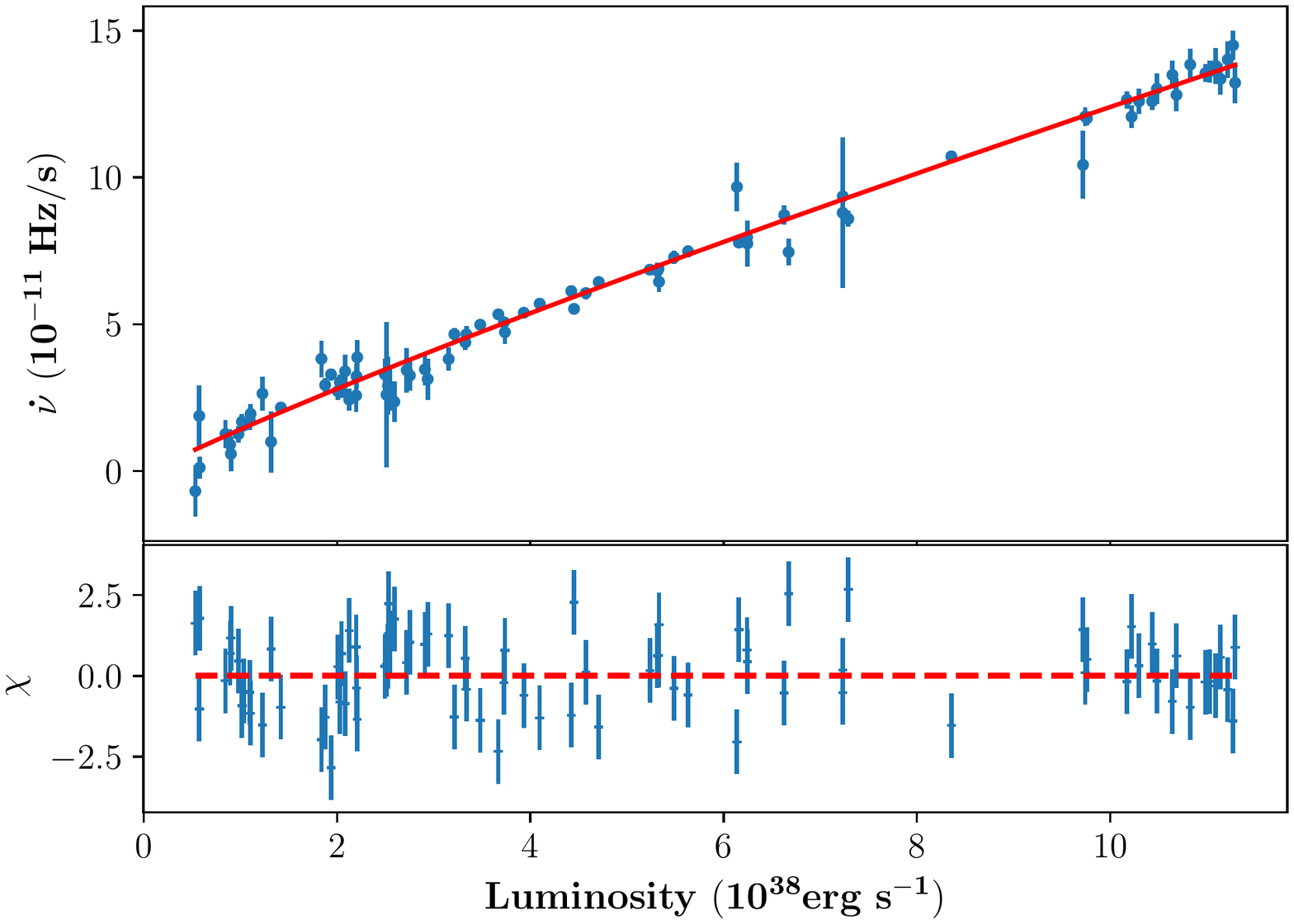}
	\caption{\footnotesize Upper panel: The luminosity vs. the frequency derivative. The red line is the fit using the GL torque model, and blue dots with 1$\sigma$ error bars are from the \textit{NICER} data. Lower panel: the residuals to the best-fit model.}
	\label{fig:fit}
\end{figure}

$k$ is usually assumed to be $0.5-1$ for a geometrically thin accretion disk. However, different values have also been proposed \citep{Bozzo2018,Monkkonen2019,Chashkina2019, Doroshenko2020,Ji2020}. We show the fitting result in Figure~\ref{fig:fit}, which indicates that the GL model can well describe the spin evolution of the source ({$\chi^2=110.7, 82\ dof$}) and the inferred magnetic field is ($5.1\pm0.2$)$\times10^{12}\,(\frac{k}{0.5})^{-7/4}$\,G. 
Using $k=0.52$ as proposed in the GL model, $B$ is then ($4.8\pm0.2)\times10^{12}$ G which is close to an order of magnitude lower than that derived from $L_{\rm crit}$.

In literature, there are several calculations on the fastness parameter under different conditions \citep{Ghosh1979,Wang1987,Wang1995,Kluzniak2007}, which however do not have a significant influence on the conclusion for a slow rotator ($\omega\sim0$) like RX J0209.6$-$7427. In comparison, the \cite{Wang1995} model has the spin-up rate as:
\begin{equation}
    \dot{\nu} = \dfrac{n(\omega)}{2\pi I}\dot{M}(GM R_{\rm m})^{1/2}   \, ,
\label{torqueWang}
\end{equation}
with $n(\omega)\approx 7/6$. If taking the same definitions of $R_{\rm A}$ and $\mu$, then the \cite{Wang1995} model will predict a magnetic field a factor of 1.8 larger than the GL model, no matter what value of $k$ is used.

Unlike the Galactic X-ray pulsars whose distances usually have large uncertainties, the distance of RX J0209.6$-$7427 is well known and with small relative uncertainty of 5\%, thus resulting in a relative uncertainty on the measured luminosity of 10\%. This will cause a relative uncertainty of 9\% on the estimated magnetic field using $L_{\rm crit}$, and of 30\% using the spin evolution approach. On the other hand, the measurement uncertainty on the luminosity from our analysis is negligible, compared to the uncertainties induced by different estimation methods and from the distance. In summary, the magnetic field strength inferred from $L_{\rm crit}$ is $(1.7-2.2)\times10^{13}$ G, and from different torque models is $(4.8-8.6)\times10^{12}$ G. Even after taking into account uncertainties of both methods and those from the distance and the measurements, we still got inconsistent magnetic field strength for RX J0209.6$-$7427.


Similar discrepancy between different approaches has also been reported for SMC X-3 and Swift J0243.6+6124. Possible existence of multipole magnetic fields was suggested for SMC X-3 \citep{Tsygankov2017} and Her X-1 \citep{Monkkonen2022}. For Swift J0243.6+6124, \cite{Doroshenko2020} proposed that either a small magnetosphere size with $k\sim 0.1-0.2$ or the presence of multipole strong magnetic field can explain the discrepancy. The recent discovery of a spin phase-dependent CRSF with high statistical significance at energies around 120$-$146 keV from Swift J0243.6+6124 has lead to a magnetic field estimate of $\sim 1.6\times 10^{13}$ G, which is an order of magnitude higher than those estimated from other methods, such as the spin evolution, but is in excellent agreement with that estimated from $L_{\rm crit}$. The authors claimed that the observed CRSF traces the multipole component of the field which dominates the field in the vicinity of the NS’s surface \citep{Kong2022} thus supports the latter scenario in \cite{Doroshenko2020}. Although the \textit{Insight}-HXMT background model uncertainty does not allow us to perform reliable spectral fit for the much fainter RX J0209.6$-$7427 at above 50 keV to investigate the possible CRSF at around 100 keV, which may have caused the dip at phase around 0.55 in the pulse profile of 80$-$130 keV band, the magnetic fields we estimated for RX J0209.6$-$7427 using the above two methods are very similar to Swift J0243.6+6124. Therefore, multipole field may also be a solution to the discrepancy between different magnetic field estimation methods for RX J0209.6$-$7427.

It is thus intriguing to ask this question: Are multipole fields common on the surfaces of NSs? Indeed, for some magnetars there are large differences between their magnetic fields measured from their observed spin-down rates and that inferred from their magnetar-unique behaviours (e.g., prolific glitches, soft gamma-ray flares, including some rare giant flares, etc.); these are commonly interpreted as due to the existence of multipole magnetic fields. Among the more distant extragalactic PULXs, a multipole strong magnetic field was suggested to interpret the properties of NGC 5907 ULX-1 \citep{Israel2017a}. Further observational and theoretical studies on the different aspects of the pulsational properties of accreting X-ray pulsars, especially those with high luminosity, are needed to fully understand the topology of the magnetic fields of these NSs.

\section{Conclusion} \label{conclude}

RX J0209.6$-$7427 is a new transient PULX in the SMC, whose ULX nature was identified during its super-Eddington outburst in 2019 November. The analysis of the broad band \textit{Insight}-HXMT data revealed its pulsed emission up to 180 keV and certainly above 130 keV, and we demonstrated that this emission is from the fan beam pattern of the accretion column. This is the highest energy pulsation detected so far from all PULXs not in the Milky Way. With the more accurately determined distance than that of the Galactic X-ray pulsars, we show that its peak luminosity of ($1.1\pm 0.06)\times 10^{39}\, \rm erg\ s^{-1}$ is intrinsic rather than highly beamed toward a certain direction. Moreover, the longer span \textit{NICER} monitoring data suggested a state transition from subcritical to supercritical accretion regime occurred at around $3\times10^{38}\, \rm erg\ s^{-1}$, which allowed us to estimate the surface magnetic field of RX J0209.6$-$7427 to be $(1.7-2.2)\times10^{13}$ G However, from its spin evolution we obtained a magnetic field about one order of magnitude lower. We interpreted the two values of magnetic fields as the dipole and multipole magnetic fields of the neutron star, similar to what has been proposed for the Galactic PULX Swift J0243.6+6124. Although there are still some detailed differences between the two closest PULXs, the striking similarity between them may suggest a common nature of their neutron stars and their ultraluminous emissions. This may shed light on better understanding of the nature of more distant extragalactic PULXs.

\acknowledgments
We thank Alexander Mushtukov for discussions on estimating the NS surface magnetic field from the observed very high critical luminosity, as well as the scenario of multipole surface magnetic field. This work made use of the data from the \textit{Insight}-HXMT mission, a project funded by the China National Space Administration (CNSA) and the Chinese Academy of Sciences (CAS). This research also made use of data obtained with the \textit{NICER}, a NASA experiment placed on the International Space Station (ISS). The \textit{Insight}-HXMT team gratefully acknowledges the support from the National Program on Key Research and Development Project (grant No. 2021YFA0718500) from the Ministry of Science and Technology of China (MOST). The authors are thankful for support from the National Natural Science Foundation of China under grants U1938103, 12041303, U1938109, U1838202, U1838201, U1838115, U1838104, 12073029, U1838107, U1938201 and U2038101. X.H. is supported by the Light of West China Program of the CAS. L.J. is supported by the Guangdong Major Project of Basic and Applied Basic Research (Grant No. 2019B030302001).

\appendix

\section{Determine the highest energy pulsation}
\label{highestE}
\begin{figure}[ht!]
	\centering
	\includegraphics[scale=0.4]{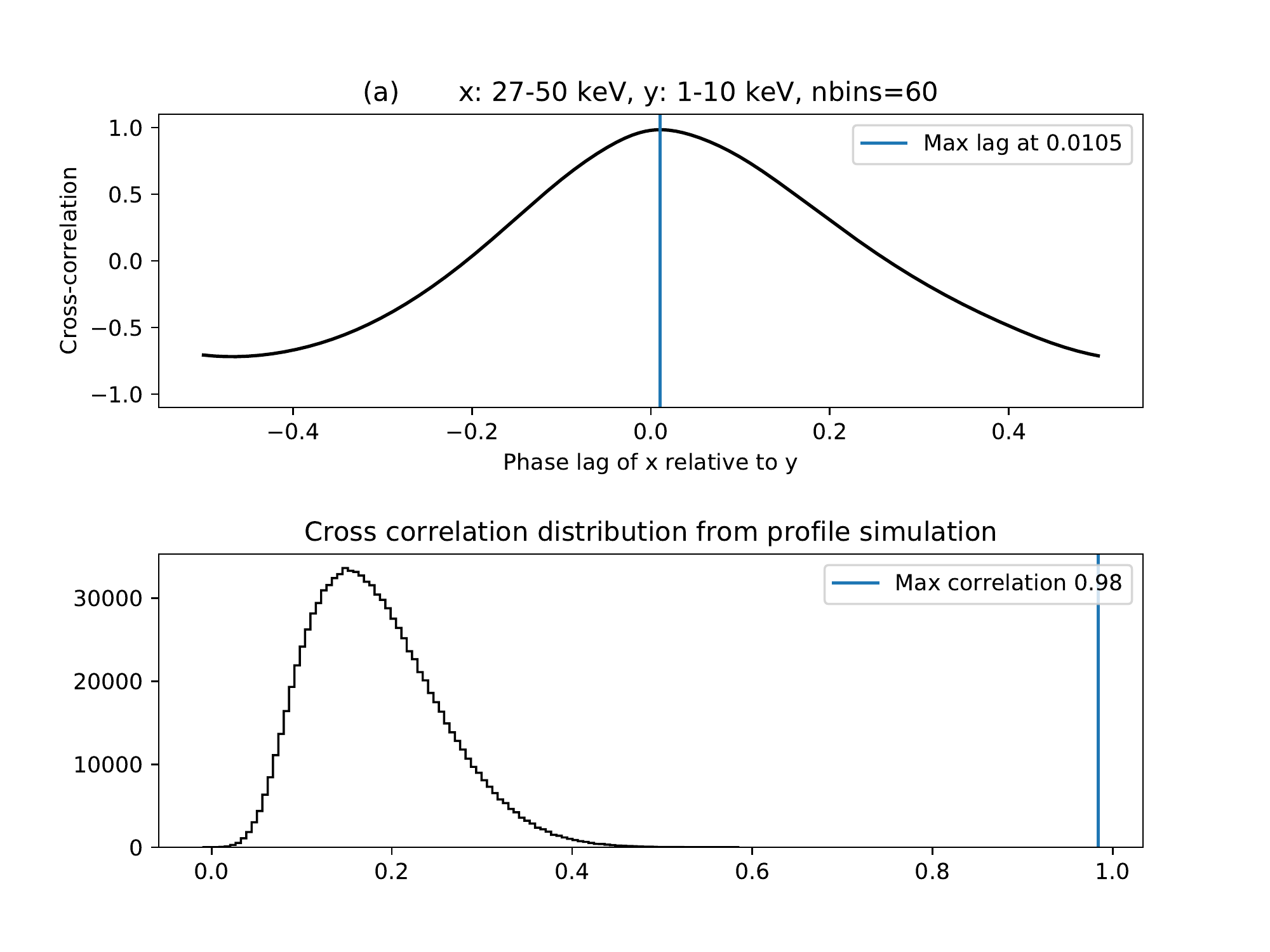}
	\includegraphics[scale=0.4]{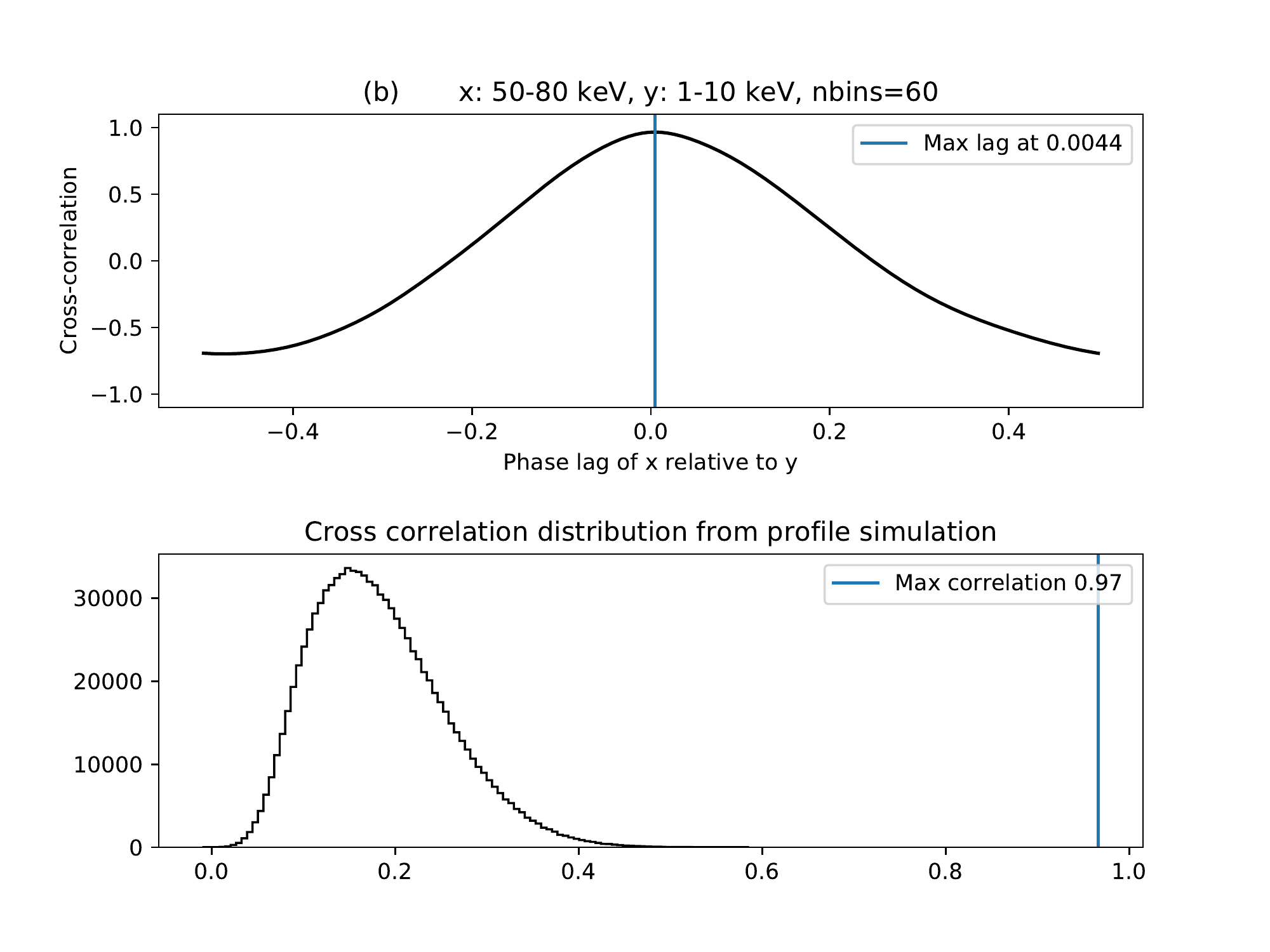}
	\includegraphics[scale=0.4]{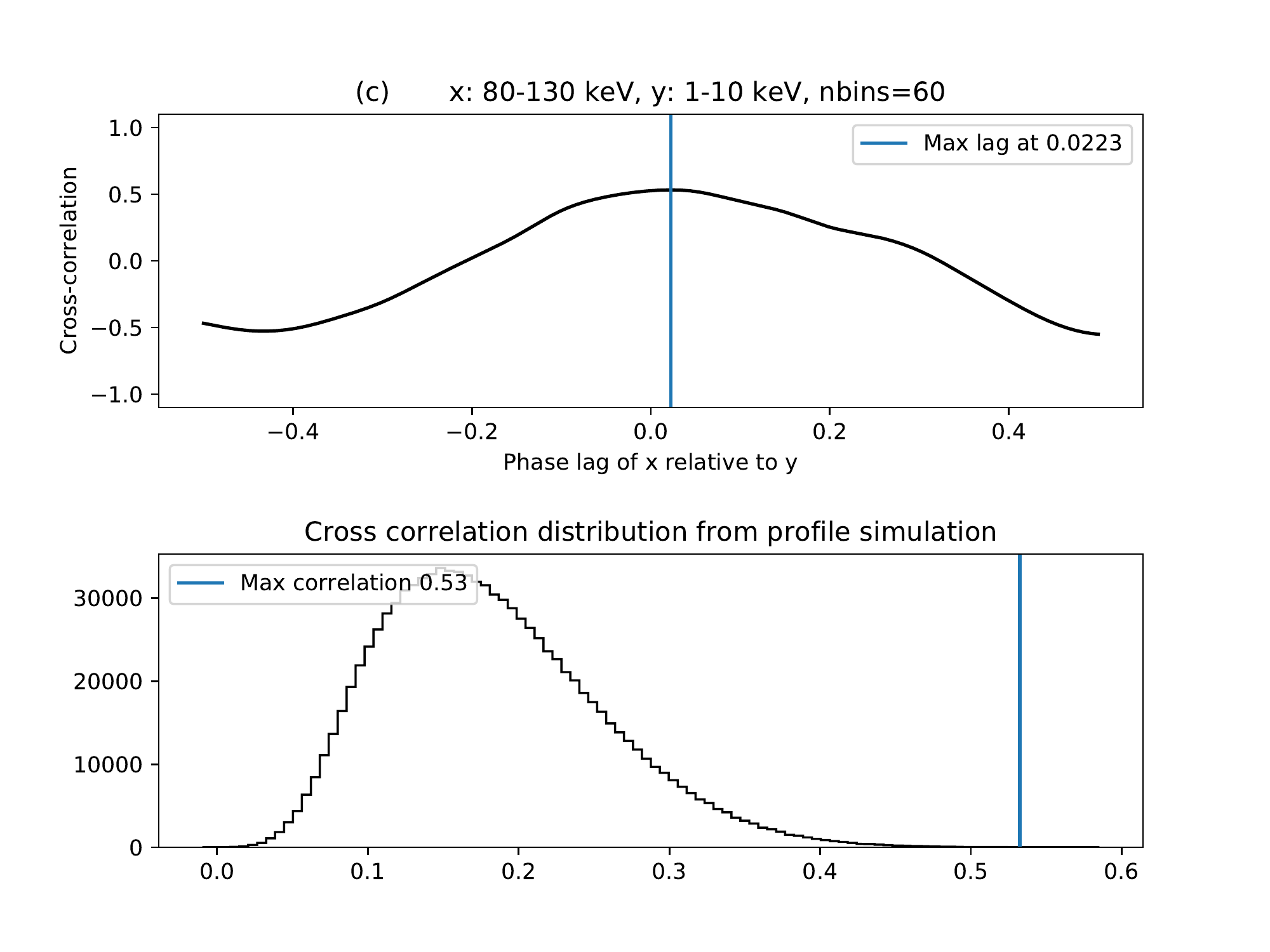}
	\includegraphics[scale=0.4]{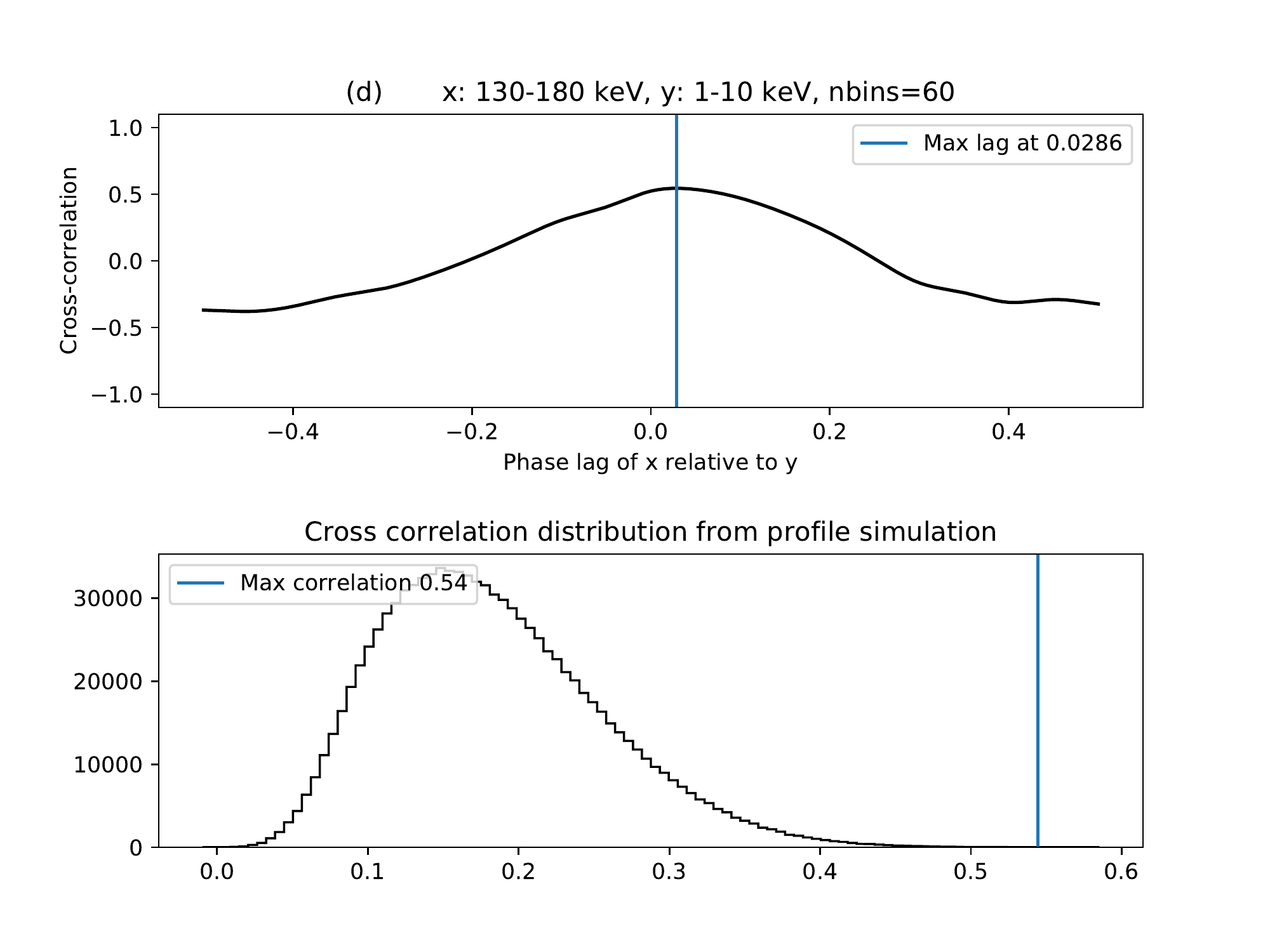}
	\caption{ \footnotesize (a), (b), (c), (d): Examples of cross correlation of different HE pulse profiles with the LE $1-10$ keV profile. In each example, the upper panel shows the cross correlation vs. phase lag for a given HE profile with the LE $1-10$ keV profile. the lower panel shows the cross correlation distribution from $10^6$ Monte-Carlo simulations of two non-correlated profiles.}
	\label{fig:simuCross}
\end{figure}

\begin{figure}[ht!]
	\centering
	\includegraphics[scale=0.6]{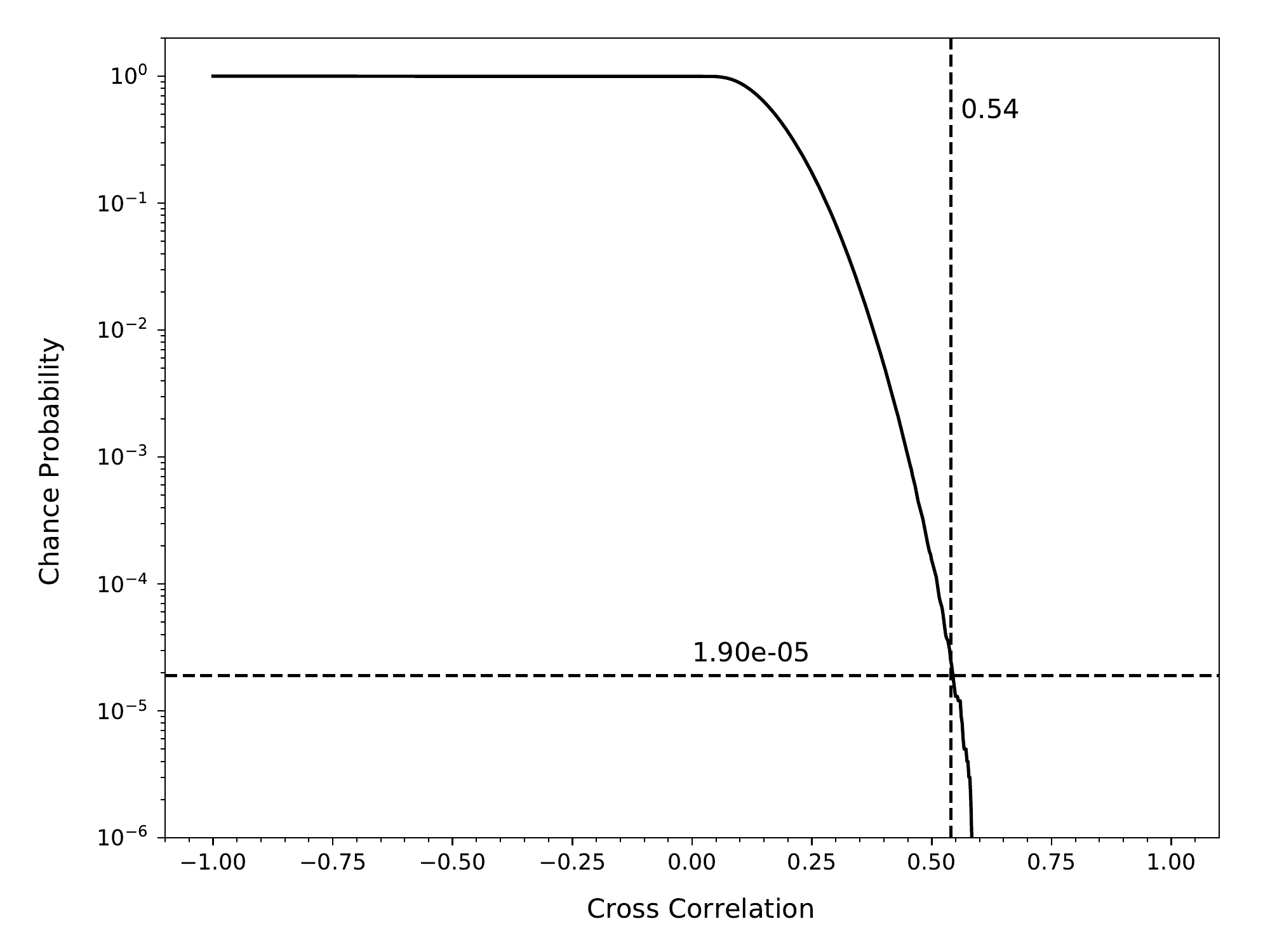}
	\caption{ \footnotesize  The cross correlation from 1000000 simulations of two non-correlated profiles and the corresponding chance probability. The dashed vertical and horizontal lines denote the cross correlation and chance probability for the HE $130-180$ keV profile, respectively. Note that the chance probability for a $3\sigma$ significance is 0.0027. The significance of the HE $130-180$ keV profile is 4.3$\sigma$.}
	\label{fig:flatproba}
\end{figure}

We apply the cross correlation technique to determine the highest energy pulsation of RX J0209.6$-$7427 using \textit{Insight}-HXMT data. We consider HE profiles in the energy range of 20$-$250 keV. We first calculate the cross correlation of a given HE profile with the 1$-$10 keV LE profile. Then we use Monte-Carlo method to simulate the cross correlation distribution between two non-correlated profiles: 
\begin{itemize}
\item The first profile is generated by performing Poisson sampling of a flat profile; the mean of the simulated profile is taken as that of each observed HE profile. 
\item The second profile is generated by performing Poisson sampling for each bin of the 1$-$10 keV LE profile. 
\item Calculate the cross correlation for the two profiles. 
\item The simulation has been repeated for $10^6$ times to obtain the distribution of the cross correlation. 
\end{itemize}
Figure~\ref{fig:simuCross} shows a few examples of the cross correlation of HE profiles and the distribution of the cross correlation from simulated profiles. To reduce the window effect, both the data and simulated profiles have been extended to 3 spin periods (nbins in Figure~\ref{fig:simuCross}). The number of bins was chosen arbitrary, but we have tested different binning of 10, 20, 30, 40, 50, 80, 100 per periods, all resulting in similar significance of the given HE pulse profile. The blue line in the plot of distribution indicates the cross correlation of a given HE profile. The significance of such cross correlation can then be evaluated by calculating the proportion of the part on the right of the blue line, i.e., the chance probability ($p$-value) that the result is obtained by chance given that the null-hypothesis (the given HE profile is flat) is true. We show the simulated cross correlations and chance probabilities in Figure~\ref{fig:flatproba}. We find that the pulsation is detected up to the 130$-$180 keV band with a significance of 4.3$\sigma$. In addition, we note that the phase lag (upper panel of each example in Figure~\ref{fig:simuCross}) of each HE profile relative to the 1$-$10 keV LE profile is close to zero, which means that the broad band pulse profiles of RX J0209.6$-$7427 are all in phase.

\section{Model of accretion column and application to RX J0209.6$-$7427}
\label{ACmodel}

We now demonstrate that the $\sim$130 keV pulsation of RX J0209.6$-$7427 can only originate from the fan beam model which does not have significant beaming effect; this conclusion is robust, since it is less dependent on the exact assumptions of the AC and NS properties. In the AC, the factor which determines whether a photon can escape from the AC is the optical depth:
\begin{equation}
    \tau = n_{\rm e} \sigma l    \, ,
\end{equation}
with $n_{\rm e}$ the electron number density, $\sigma$ the scattering cross section between photons and electrons and $l$ the photon propagation length. The scattering cross section depends on the photon energy, the magnetic field, the polarization state and the photon direction momentum. We are interested in photons propagating parallelly and perpendicularly to the magnetic field. We have \citep{Arons1987}:
\begin{enumerate}
 \item $E<E_{\rm c}$
 
   \begin{itemize}
     \item  O-Mode 
      \begin{equation}
       \begin{aligned}
          \sigma_{\parallel} =  \sigma_{\rm T} (E/E_{\rm c})^2      \, ,  \\
          \sigma_{\perp}  = \sigma_{\rm T}      \, ,
       \end{aligned}
      \end{equation}
     \item  X-Mode 
       \begin{equation} 
        \sigma_{\rm X} =  \sigma_{\rm T} (E/E_{\rm c})^2      \, ,
        \end{equation}
      \end{itemize}

 \item $E \geq E_{\rm c}$
        \begin{equation} 
        \sigma_{\rm O}  = \sigma_{\rm X} =  \sigma_{\rm T}       \, ,
        \end{equation}
        
\end{enumerate}
where O-mode is the ordinary polarization mode with the electric vector in the plane of the magnetic field $\boldsymbol{B}$ and the photon momentum $ \boldsymbol{\hbar k}$, X-mode is the extraordinary polarization mode with the electric vector perpendicular to the plane of ($\boldsymbol{B}, \boldsymbol{k}$), $\sigma_{\rm T} $ is the Thomson scattering cross section, and $E_{\rm c}=11.6B_{12}/(1+z)$ is the CRSF line energy. $B_{12}$ is the magnetic field in units of $10^{12}$ G and $z$ is the gravitational redshift, usually between 0 and 0.3 in the case of NSs. 

For $n_{\rm e}$ and $l$, we make use of the analytical expressions in \cite{Becker2012}. In the supercritical high luminosity case, a radiation-dominated shock is formed above the NS whose height increases with increasing luminosity. $n_{\rm e}$ is calculated from the mass conservation relation (Eq. 12) and by assuming that the inflow velocity in the AC equals to the post-shock velocity (Eq. 4) which is $1/7$ of the free-fall velocity of the accreting matter approaching the top of the radiation-dominated shock, and that all the matter are accreted on to the NS ($L_{X} = GM\dot{M}/R$). The photon propagation length $l$ in the direction parallel to the magnetic field can be estimated by the height of the radiation-dominated shock $H$ (Eq. 16). In the direction perpendicular to the magnetic field, $l$ is estimated by the AC radius $r_{0}$ (Eq. 23). In our calculations, we took typical values for involved parameters as shown in \cite{Becker2012}: $M=1.4\,M_{\rm \odot}$, $R=10$ km, $\xi = 0.01$, $\Lambda=0.1$ and $\tau_{*}=20$.

The critical luminosity can be calculated as $L_{\rm crit} = 1.49\times 10^{37} B_{12}^{16/15}$. We consider different magnetic field strengths of $10^{11}$, $10^{12}$ and $10^{13}$ G and assuming $z=0$ (we verified that taking $z=0.3$ does not affect our conclusion). The corresponding $L_{\rm crit}$ is then $1.5\times 10^{36}$, $1.5\times 10^{37}$ and $1.5\times 10^{38} \, \rm erg \, s^{-1}$, respectively. From our analysis, RX J0209.6$-$7427 is in the supercritical state before 58880 MJD for all three values of the magnetic field, thus it is appropriate to apply the equations aforementioned to calculate the optical depth. We consider the case of peak luminosity of $10^{39} \, \rm erg \, s^{-1}$. The results are shown in Table~\ref{table:optdepth}.

\begin{figure}[ht!]
	\centering
	\includegraphics[scale=0.2]{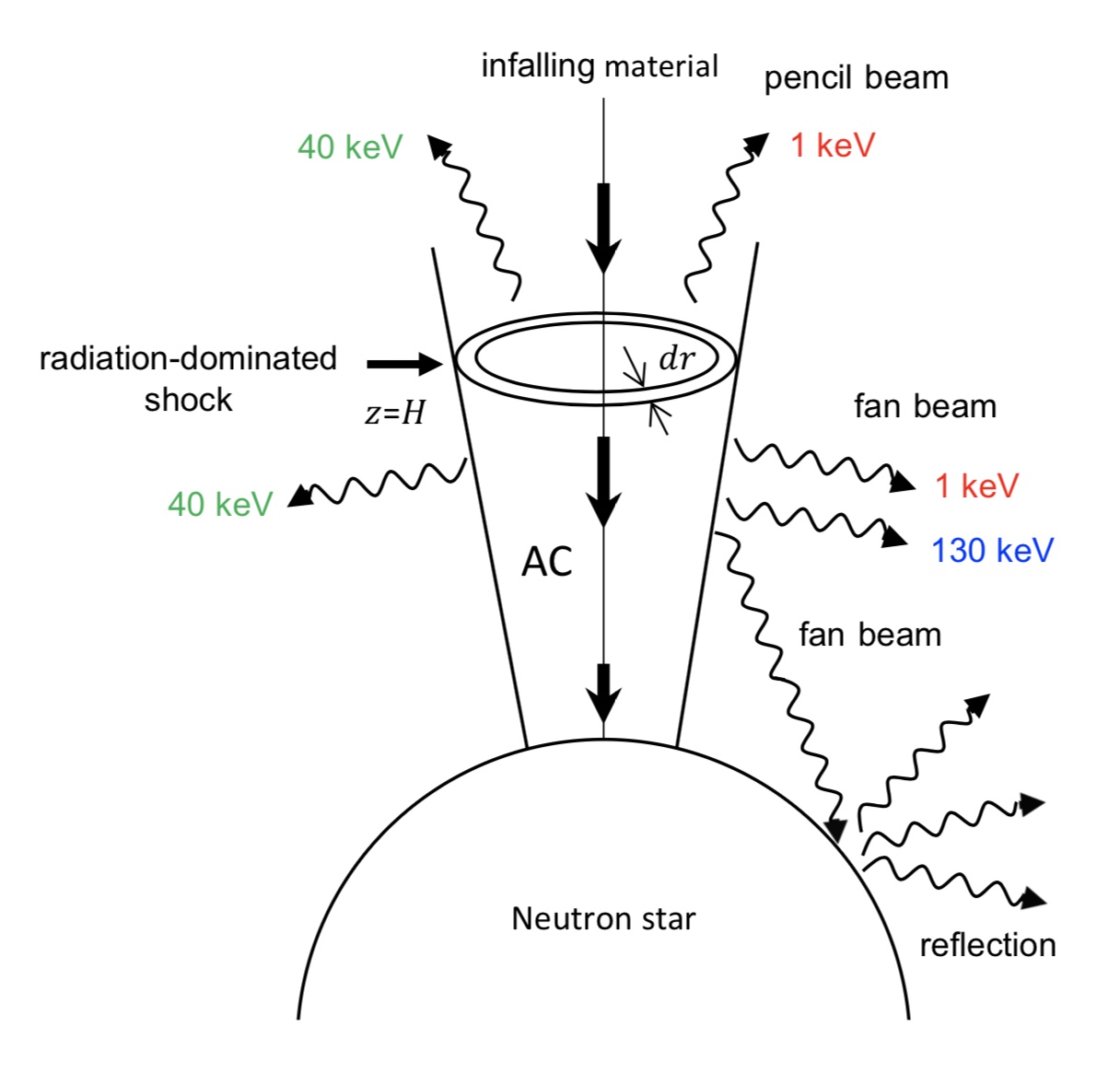}
	\caption{\footnotesize Illustration of the AC and the emission region of photons with different energies. Low and intermediate energy photons (e.g., 1$-$40 keV) may be emitted in both pencil beam and the upper fan beam, but high energy photons (e.g., from about 50 to above 130 keV) can only be emitted in the lower fan beam and also reflected from the surface of the neutron star; the reflected flux should have non or very low spin modulation and thus reduces the PF in the highest energy bands. $H$ is the height of the radiation-dominated shock and $dr$ is the thickness of the radiation shell of the fan beam. See \cite{Basko1976} for detailed description of the AC.} 
	\label{fig:geometry}
\end{figure}

Three conclusions can be drawn from the results in the case of supercritical high luminosity state (Figure~\ref{fig:geometry}): (1) For photons of 130 keV, no matter what kind of polarization mode they have, the optical depth in the direction parallel or perpendicular to the magnetic field is always $\gg1$, thus they can not escape directly from the core of the AC. However, it is more appropriate to consider a thin shell in the direction perpendicular to the magnetic field, i.e., the wall of the AC. Setting $\tau_{\perp,\rm wall}=1$ then gives the shell thickness $dr$ for photons to be able to just escape from the wall.  We can see that $dr$ is far less than $H$ in the direction parallel to the magnetic field and the AC radius $r_{0}$. So the photons of 130 keV can, and can only escape from the wall of the AC, forming a fan beam. 

(2) For photons of 1 keV, if it is O-mode, then they can escape from the wall of the AC, similar to the 130 keV photons since $\tau_{\perp}$ is independent of photon energy. If it is X-mode, in the case of a strong magnetic field as $10^{13}$ G, the cross section in the direction perpendicular to the magnetic field is largely reduced, so that the optical depth $\tau_{\perp,\rm X}\ll 1$ and the photons can also escape without problem. This can be inferred as well from the shell thickness $dr$ which in this case is already larger than the column radius $r_{0}$. This  implies that the AC is optically thin and photons from the center of the column can escape. So the 1 keV photons can form a fan beam as the 130 keV photons do. On the other hand, the optical depth $\tau_{\parallel}$ for O-mode photons is still larger than unity with a strong magnetic field of $10^{13}$ G, so the photons can not escape. Nevertheless, $\tau_{\parallel}$ can be reduced to less than unity if the magnetic field is increased by two times. Besides, the ratio of $\tau_{\parallel}$ to $\tau_{\perp}$ is $\ll 1$ for $B\sim 10^{12-13}$ G, suggesting that photons will escape more easily from the direction parallel to the magnetic field. In this case, the 1 keV photons can also escape from the top of the AC along the magnetic field lines in a pencil beam pattern. 

(3) For photons with intermediate energy like 40 keV, they can escape from the wall of the AC as explained for the 130 keV and 1 keV photons to form a fan beam. However, for O-mode photons, $\tau_{\parallel}$ is comparable to $\tau_{\perp}$ with a ratio of 7 for a magnetic field of $10^{13}$ G. In this case, pencil beam can not be excluded. Considering the simplification of the model, low and intermediate energy photons probably can also escape from the pencil beam for $B\sim 10^{12-13}$ G even at $L\sim 10^{39}\, \rm erg\ s^{-1}$, though with significant attenuation.

Before 58880 MJD when RX J0209.6$-$7427 was in the supercritical luminosity state, its pulse profile exhibits one main peak and one minor peak at lower energies separated by $\sim$0.5 in phase, while at above 27 keV, only the main peak is prominent reaching up to 180 keV and certainly above 130 keV. We have demonstrated that the 130 keV photons can only originate from the fan beam, therefore the main peak is fan beam in origin. It is not surprising that the main peak is also visible at lower energies, since we have demonstrated that both 130 keV and 1 keV photons can form a fan beam. 


\bibliography{RXJ0209.bib}{}
\bibliographystyle{aasjournal}


\begin{table}[ht!]
\small
\caption{The orbit and spin parameters of RX J0209.6$-$7427.}
\label{table:lcfit}
\begin{center}
\begin{tabular}{l l l  c c}
\hline \hline
Parameters  & Value  \\
\hline
$P_{\rm orb}$ (day)   &  $47.97\pm0.16$  \\
$a\rm sin$ $i$ (light-sec)   &  $168\pm3$ \\
$e$ &  $0.317\pm0.007$ \\
$\omega$     & $-98.5\pm3.8$ \\
$T_{\omega}$ (MJD) & $58782.7\pm0.2$\\
\hline
PEPOCH (MJD)   &  $58800$\\
$\nu_{0}\, \rm{(Hz)} $  &  $0.107557(8)$  \\
$\nu_{1}\, (10^{-10}\, \rm Hz\, s^{-1}$)   &  $-1.66\pm0.15$  \\
$\nu_{2}\, (10^{-16}\, \rm Hz\, s^{-2}$)   &  $3.14\pm0.20$  \\
$\nu_{3}\, (10^{-22}\, \rm Hz\, s^{-3}$)   &  $-2.25\pm0.18$  \\
$\nu_{4}\, (10^{-29}\, \rm Hz\, s^{-4}$)   &  $1.01\pm0.11$  \\
$\nu_{5}\, (10^{-35}\, \rm Hz\, s^{-5}$)   &  $-2.68\pm0.38$  \\
$\nu_{6}\, (10^{-42}\, \rm Hz\, s^{-6}$)   &  $3.28\pm0.65$  \\
\hline
$\chi^{2}$ & 0.992(76)\\
\hline
\end{tabular}
\end{center}
\end{table}


\begin{table}[ht!]
 \scriptsize
\begin{center}
\begin{threeparttable}
\caption{\insight{} best-fit spectral results for the 5-observation combined phase-averaged spectrum.} 
\label{table:fitave}
\begin{tabular}{lcc}
 \toprule
 Component  & Parameters &   \insight{} background model  \\[0.8ex]
 \midrule
Tbabs          &   $N_{\rm H} \, (10^{22} \, \rm cm^{-2})$            & $0.158$  \\[0.8ex]
cufoffpl       &   $\Gamma$                                           & $0.72^{+0.01}_{-0.01}$   \\[0.8ex]
               &   $E_{\rm cut}$ (keV)                                & $11.38^{+0.12}_{-0.11}$   \\[0.8ex]
               &   norm ($10^{-2}$)                                   &     $9.74^{+0.09}_{-0.09}$   \\[0.8ex]
 
 bbodyrad      & $kT$ (keV)                                           & $0.18^{+0.01}_{-0.01}$   \\[0.8ex]
               & norm                                                 & $18400^{+3810}_{-3050}$   \\[0.8ex]
              & $R_{\rm BB}$ (km)$^{a}$                               & $746^{+340}_{-304}$   \\[0.8ex]

Gaussian      & $E_{\rm Fe}$ (keV)                                    & $6.58^{+0.03}_{-0.03}$     \\[0.8ex]
              &$\sigma_{\rm Fe}$  (keV)                               & $0.24^{+0.05}_{-0.05}$        \\[0.8ex]
              & norm ($10^{-3}\, \rm cm^{-2} \,  s^{-1}$)             & $1.11^{+0.15}_{-0.14}$       \\[0.8ex]  
\hline              
              & $\chi^{2}/d.o.f$                                       & $1464/1323$   \\
             & $F_{\rm  X}\,(10^{-9} \, \rm cm^{-2} \,  s^{-1})$      & $3.06^{+0.02}_{-0.02}$ \\[0.8ex]
              & $L_{\rm  X}\,(10^{38} \, \rm erg \, s^{-1})^{b}$       & $11.08^{+0.06}_{-0.06}$  \\

\bottomrule
\end{tabular}
\begin{tablenotes}
     \scriptsize
      \item{}{$^{a}$ The black body component radius was estimated from the normalization of the model assuming a distance to the SMC of 55 kpc (i.e., $D_{10}=5.5$).}
      \item{}{$^{b}$ Luminosity was calculated in the energy band of 1$-$150 keV.}
      \vspace{0.5cm} 
     \item {}{\textit{Notes.} The fits are performed in the energy bands of $1-9$ keV (LE), $8-30$ keV (ME) and $28-50$ keV (HE). The best-fit model is \texttt{Tbabs*(cutoffpl+bbodyrad+Gaussian)}. All the errors are given at the 68\% confidence level.}
\end{tablenotes}
\end{threeparttable}
\end{center}
\end{table}

\begin{table}[ht!]
 \scriptsize
\begin{center}
\begin{threeparttable}
\caption{\insight{} best-fit spectral results for the 5-observation combined main peak. } 
\label{table:fitbigpeak}
\begin{tabular}{lccc}
 \toprule
 Component  & Parameters &  OFF spectrum as background  &  \insight{} background model  \\[0.8ex]
 

\hline  
Tbabs          &   $N_{\rm H} \, (10^{22} \, \rm cm^{-2})$     & $0.158$                   & $0.158$  \\[0.8ex]
cufoffpl       &   $\Gamma$                                    & $0.64^{+0.03}_{-0.03}$    & $0.71^{+0.01}_{-0.01}$   \\[0.8ex]
               &   $E_{\rm cut}$ (keV)                         & $12.48^{+0.39}_{-0.37}$   & $11.70^{+0.16}_{-0.14}$   \\[0.8ex]
               &   norm ($10^{-2}$)                            & $5.78^{+0.16}_{-0.16}$    & $11.24^{+0.08}_{-0.09}$   \\[0.8ex]
 
 bbodyrad      & $kT$ (keV)                                    & $2.17^{+0.48}_{-0.36}$    & $1.47^{+1.80}_{-1.16}$   \\[0.8ex]
               & norm                                          & $0.22^{+0.25}_{-0.15}$    & $0.36^{+0.74}_{-0.07}$   \\[0.8ex]
               & $R_{\rm BB}$ (km)$^{a}$                       & $2.6^{+2.7}_{-2.1}$       & $3.3^{+4.7}_{-1.4}$   \\[0.8ex]

 bbodyrad      & $kT$ (keV)                                    & ...                        & $0.16^{+0.01}_{-0.01}$      \\[0.8ex]
               & norm                                          & ...                        & $28500^{+9090}_{-6600}$   \\[0.8ex]
               & $R_{\rm BB}$ (km)$^{a}$                       & ...                        & $929^{+524}_{-447}$      \\[0.8ex]

Gaussian      & $E_{\rm Fe}$ (keV)                             & ...                       & $6.61^{+0.04}_{-0.04}$     \\[0.8ex]
              & $\sigma_{\rm Fe}$  (keV)                       & ...                       & $0.21^{+0.06}_{-0.05}$        \\[0.8ex]
              & norm ($10^{-3}\, \rm cm^{-2} \, \rm s^{-1}$)   & ...                       & $1.14^{+0.13}_{-0.13}$       \\[0.8ex]  
\hline              
              & $\chi^{2}/d.o.f$                                      & $1461/1395$              &$1477/1325$   \\
              & $F_{\rm  X}\,(10^{-9}\, \rm cm^{-2} \,s^{-1})$        & $2.51^{+0.04}_{-0.09}$   & $3.68^{+0.02}_{-0.02}$ \\[0.8ex]
              & $L_{\rm  X}\,(10^{38}\, \rm erg \, s^{-1})^{b}$      & $9.08^{+0.15}_{-0.32}$   & $13.31^{+0.08}_{-0.09}$ \\

\bottomrule
\end{tabular}
\begin{tablenotes}
     \scriptsize
      \item{}{$^{a}$ The black body component radius is estimated from the normalization of the model assuming a distance to the SMC of 55 kpc (i.e., $D_{10}=5.5$).}
      \item{}{$^{b}$ Luminosity is calculated in the energy band of 1$-$150 keV.}
      \vspace{0.5cm} 
      \item{}{\textit{Notes.} The fits are performed in the energy bands of $1-9$ keV (LE), $8-30$ keV (ME) and $28-150$ keV (HE) when using OFF spectrum as the background model, and in the energy bands of $1-9$ keV (LE), $8-30$ keV (ME) and $28-50$ keV (HE) when using the \insight{} background model. The best-fit models for the 5-observation combined spectra using the OFF spectrum and \insight{} background model are \texttt{Tbabs*(cutoffpl+bbodyrad)} and \texttt{Tbabs*(cutoffpl+bbodyrad+bbodyrad+Gaussian)}, respectively. All the errors are given at the 68\% confidence level.}
\end{tablenotes}
\end{threeparttable}
\end{center}
\end{table}

\begin{table}[ht!]
 \scriptsize
\begin{center}
\begin{threeparttable}
\caption{\insight{} best-fit spectral results for the 5-observation combined minor peak.} 
\label{table:fitsmallpeak}
\begin{tabular}{lccc}
 \toprule
 Component  & Parameters &  OFF spectrum as background   &  \insight{} background model  \\[0.8ex]
\hline  
Tbabs          &   $N_{\rm H} \, (10^{22} \, \rm cm^{-2})$     & $0.158$                    & $0.158$  \\[0.8ex]
cufoffpl       &   $\Gamma$                                    & $1.15^{+0.09}_{-0.10}$     & $0.80^{+0.02}_{-0.02}$   \\[0.8ex]
               &   $E_{\rm cut}$ (keV)                         & $22.17^{+12.95}_{-6.29}$   & $10.86^{+0.30}_{-0.28}$   \\[0.8ex]
               &   norm ($10^{-2}$)                            & $2.78^{+0.21}_{-0.20}$     & $7.42^{+0.19}_{-0.19}$   \\[0.8ex]
 
 bbodyrad     & $kT$ (keV)                                    &...                         & $0.19^{+0.01}_{-0.01}$   \\[0.8ex]
               & norm                                         &...                         & $12400^{+3820}_{-2740}$   \\[0.8ex]
               & $R_{\rm BB}$ (km)$^{a}$                     &...                          & $612^{+340}_{-288}$     \\[0.8ex]
               
Gaussian      & $E_{\rm Fe}$ (keV)                             & ...                       & $6.48^{+0.07}_{-0.07}$     \\[0.8ex]
              &$\sigma_{\rm Fe}$  (keV)                        & ...                       & $0.33^{+0.09}_{-0.07}$      \\[0.8ex]
              & norm ($10^{-3}\, \rm cm^{-2} \, s^{-1}$)       & ...                       & $1.17^{+0.27}_{-0.24}$      \\[0.8ex]  
\hline              
              & $\chi^{2}/d.o.f$                                       & $1380/1314$  &$1362/1309$   \\
              & $F_{\rm  X}\,(10^{-9} \, \rm cm^{-2} \, s^{-1})$       & $0.61^{+0.08}_{-0.14}$   & $1.84^{+0.02}_{-0.03}$ \\[0.8ex]
              & $L_{\rm  X}\,(10^{38} \, \rm erg \, s^{-1})^{b}$       & $2.21^{+0.29}_{-0.51}$   & $6.67^{+0.06}_{-0.10}$ \\

\bottomrule
\end{tabular}
\begin{tablenotes}
     \scriptsize
      \item{}{$^{a}$ The black body component radius is estimated from the normalization of the model assuming a distance to the SMC of 55 kpc (i.e., $D_{10}=5.5$).}
      \item{}{$^{b}$ Luminosity is calculated in the energy band of 1$-$150 keV.}
      \vspace{0.5cm} 
      \item{}{\textit{Notes.} The fits are performed in the energy bands of $1-9$ keV (LE) and $8-30$ keV (ME). The best-fit models for the 5-observation combined spectra using the OFF spectrum and \insight{} background model are \texttt{Tbabs*cutoffpl} and \texttt{Tbabs*(cutoffpl+bbodyrad+Gaussian)}, respectively. All the errors are given at the 68\% confidence level.}
\end{tablenotes}
\end{threeparttable}
\end{center}
\end{table}


\begin{table}[ht!]
\begin{center}
\begin{threeparttable}
\caption{Optical depth calculation for different polarization modes, magnetic fields and photon energies. } 
\label{table:optdepth}
\begin{tabular}{lccccccc}
 \toprule
 \hline
 $E \ \rm (keV)$  & $B_{12}$ & $\tau_{\parallel}$  & $\tau_{\perp}$  &$\tau_{\parallel}/\tau_{\perp}$  &$r_{0} \ \rm (cm)$  & $dr \ \rm (cm)$  & $dr/r_{0}$ \\
 \midrule
\multicolumn{8}{c}{O-mode}      \\
\hline
$130$       & $0.1$      & $5.5\times10^{3}$        &  $3.4\times10^{2}$      & $16$         & $7\times10^{5}$       & $2\times10^{3}$     &$0.003$  \\
             & $1$        & $2\times10^{4}$       & $6.6\times10^{2}$       & $30$        & $4\times10^{5}$       & $6\times10^{2}$     &$0.002$  \\
             & $10$       &$7.6\times10^{4}$        & $1.3\times10^{3}$       & $58$         & $2\times10^{5}$       & $2\times10^{2}$     &$0.001$  \\
$40$        & $0.1$       & $5.5\times10^{3}$       &  $3.4\times10^{2}$      & $16$         & $7\times10^{5}$       & $2\times10^{3}$     &$0.003$  \\
            & $1$         & $2\times10^{4}$       & $6.6\times10^{2}$       & $30$         & $4\times10^{5}$       & $6\times10^{2}$     &$0.002$  \\
           & $10$        &$9\times10^{3}$        & $1.3\times10^{3}$       & $7$           & $2\times10^{5}$       & $2\times10^{2}$     &$0.001$  \\               
$1$         & $0.1$       & $4\times10^{3}$       &  $3.4\times10^{2}$      & $12$         & $7\times10^{5}$       & $2\times10^{3}$     &$0.003$  \\
           & $1$         &  $1.5\times10^{2}$      &$6.6\times10^{2}$        &  $0.2$        & $4\times10^{5}$       & $6\times10^{2}$     &$0.002$   \\
           & $10$       & $5.6$                    & $1.3\times10^{3}$       & $0.004$      & $2\times10^{5}$       & $2\times10^{2}$     &$0.001$   \\
\hline
\multicolumn{8}{c}{X-mode}      \\
\hline
$130$         & $0.1$   & $5.5\times10^{3}$		 & $3.4\times10^{2}$       & $16$        & $7\times10^{5}$          & $2\times10^{3}$       &$0.003$   \\
               & $1$    & $2\times10^{4}$		 & $6.6\times10^{2}$       & $30$        & $4\times10^{5}$          &  $6\times10^{2}$      &$0.002$   \\
               & $10$   & $7.6\times10^{4}$	     & $1.3\times10^{3}$       & $58$        & $2\times10^{5}$          & $2\times10^{2}$       &$0.001$    \\
$40$           & $0.1$  & $5.5\times10^{3}$		 & $3.4\times10^{2}$       & $16$        & $7\times10^{5}$          & $2\times10^{3}$       &$0.003$    \\
               & $1$    & $2\times10^{4}$		 & $6.6\times10^{2}$       & $30$        & $4\times10^{5}$          &  $6\times10^{2}$      &$0.002$    \\
               & $10$   & $9\times10^{3}$	     & $1.5\times10^{2}$       & $60$        & $2\times10^{5}$          & $1\times10^{3}$       &$0.007$    \\
$1$            & $0.1$  & $4.1\times10^{3}$	     & $2.6\times10^{2}$       & $16$        & $7\times10^{5}$         & $3\times10^{3}$        &$0.004$    \\
               & $1$    & $1.5\times10^{2}$		         & $5$                   & $30$        & $4\times10^{5}$         & $8\times10^{4}$        &$0.203$    \\
               & $10$   & $5.6$	                 & $0.1$                 & $56$        & $2\times10^{5}$         & $2\times10^{6}$        &$10.492$   \\
 \bottomrule

\end{tabular}
\begin{tablenotes}
     \scriptsize
     \item \textit{Notes.} The photon propagation length $H$ in the direction parallel to the magnetic field lines depends only on luminosity and is $1\times10^{7}$ cm for $10^{39} \, \rm erg \, s^{-1}$ considered here.
\end{tablenotes}
\end{threeparttable}
\end{center}
\end{table}

\end{document}